\documentclass[12pt,preprint]{emulateapj}  
\usepackage{epsfig}         	

\newcommand{\eten}[1]{\mbox{$10^{#1}$}}
\newcommand{\ee}[1]{\mbox{${} \times 10^{#1}$}}
\newcommand{\msun}{\mbox{$M_\odot$}}
\newcommand{\lsun}{\mbox{$L_\odot$}}

\newcommand{\tff}{t_{\rm{ff}}}

\newcommand{\td}{\mbox{$T_d$}}

\shorttitle{Dust Heating}
\shortauthors{Urban et al.}         
\begin{document}

\title{Fragmentation and Evolution of Molecular Clouds. II: The Effect of Dust Heating}

\author{Andrea Urban\altaffilmark{1,}\altaffilmark{2},
Hugo Martel\altaffilmark{3,}\altaffilmark{4}, and Neal J. Evans II\altaffilmark{1}}

\altaffiltext{1}{Department of Astronomy, University of Texas, Austin, 
                 TX 78712}

\altaffiltext{2}{Jet Propulsion Laboratory, California Institute of Technology, Pasadena, CA 91109}

\altaffiltext{3}{D\'epartement de physique, g\'enie physique et optique,
Universit\'e Laval, Qu\'ebec, QC, G1K 7P4, Canada}

\altaffiltext{4}{Centre de Recherche en Astrophysique du Qu\'ebec}

\slugcomment{{\sc Accepted to ApJ}} 

\begin{abstract}
We investigate the effect of heating by luminosity sources in a simulation of clustered star formation.  Our heating method involves a simplified continuum radiative transfer method that calculates the dust temperature.  The gas temperature is set by the dust temperature.  
We present the results of four simulations, two simulations assume an isothermal equation of state and the two other simulations include dust heating.  We investigate two mass regimes, i.e., $84 \msun$ and $671 \msun$, using these two different energetics algorithms.
 The mass functions for the isothermal simulations and simulations which include dust heating are drastically different.  In the isothermal simulation, we do not form any objects with masses above $1\msun$.  However, the simulation with dust heating, while missing some of the low-mass objects, forms high-mass objects ($\sim 20 \msun$) which have a distribution similar to the Salpeter IMF.  
 The envelope density profiles around the stars formed in our simulation match observed values around isolated, low-mass star-forming cores.   
We find the accretion rates to be highly variable and, on average, increasing with final stellar mass.   By including radiative feedback from stars in a cluster-scale simulation, we have determined that it is a very important effect which drastically affects the mass function and yields important insights into the formation of massive stars.

\end{abstract}

\section{Introduction}\label{sec:intro}

One of the great puzzles in the field of star formation is the universality of the initial mass function (IMF) in the Galactic environment (see \citealt{salpeter}; \citealt{ms79}; \citealt{scalo}; \citealt{kroupa}; \citealt{chabrier}), specifically the slope at high mass and the characteristic turnover mass.  One of the proposed explanations for the IMF's universality is similar thermal properties of the gas in the Galaxy in star-forming regions.  The equation of state of the gas set by the thermal physics determines the level of fragmentation in a cloud and thus the mass function \citep{larson}.

The initial state of the gas that is likely to form stars is typically assumed to be dense, molecular gas which is isothermal, maintaining a temperature of $\sim$10K.  Simulations with isothermal equations of state have been used in an attempt to recreate the IMF (\citealt{klessen}; \citealt{mes06}, hereafter MES06).  These attempts have met with limited success due to the behavior of the Jeans mass, the basic unit of star formation, under the conditions of an isothermal equation of state.  The Jeans mass is proportional to $ T^{3/2} n^{-1/2}$, where $T$ is the gas temperature (K) and $n$ is the gas number density (cm$^{-3}$).  For a constant temperature, as in the isothermal case, the Jeans mass decreases as the density increases, leading to perpetual fragmentation for a collapsing
gas cloud.  In reality, the fragmentation is believed to stop when the gas becomes optically thick as the density increases.  This prevents radiation from escaping and cooling the gas back to the isothermal temperature.

In order to model this effect in simulations, a polytropic equation of state has been assumed, i.e. $P\propto \rho^{\gamma}$.  At low densities, $\gamma \le1$ and at higher densities, $\gamma > 1$.  The density at which this transition occurs and the values of $\gamma$ have been studied by many groups (see \citealt{bbb}; \citealt{li}; \citealt{bate05}; \citealt{jappsen}; \citealt{larson};  \citealt{bcb}; \citealt{clark}).

However, all of the studies mentioned above ignore a key ingredient in the attempt to re-create the IMF.  
Their assumption that the equation of state depends solely on density is likely to be valid only when the first gas cores condense out of the cloud and begin to collapse.  Once these cores begin to generate their own energy via collapse, accretion, and deuterium/hydrogen burning, a spatially-uniform equation of state is no longer valid.  
The newly formed stars will heat the gas in the cloud and influence subsequent, nearby star formation.

In order to include the energy from the forming stars in simulations, radiative transfer must be used.  
\citet{krumholz} have run simulations in which 
they include the radiative energy from young stars.  
They include this effect in order to understand the formation of massive stars.  
Therefore, they study a small size scale (collapse from $\sim 0.1$ pc to $\sim 10$ AU scales) 
at very high densities ($n\sim 10^9$cm$^{-3}$) and form $<10$ stars.  
They use the method of flux-limited diffusion (FLD) to calculate the effect of radiation on the surrounding dust.  Then they assume the dust and gas are well-coupled to calculate the gas temperature.  (This assumption is only valid in the high density regions that they study \citep{gk}).  
As discussed in their paper, \citet{krumholz} assume gray radiation which will underestimate the true dust temperature.  
Also, their method of FLD is only accurate in very optically thick regimes.  
Despite these limitations, their method of radiative transfer is suitable for their study of the formation of  massive stars.

 \citet{bate09r} also includes FLD in his simulations.  He forms less than 20 stars in his simulations which follow the collapse from scales of $\sim 0.1$pc to 0.5 AU and the largest object formed has $M< 2$\msun.  Because of the low mass of the objects formed in his simulations, he claims that it is valid to ignore the intrinsic stellar luminosity in his calculation of heating.  He only considers energy generation via work on the gas.  This only accounts for any accretion luminosity generated outside of the accretion radius.  However, since most of the accretion luminosity is released when material accretes onto the star, his calculation of the accretion luminosity is an extreme underestimate of the total value.  Hence, his results only demonstrate the minimum possible effect of radiative feedback.

Recent work by \citet{offner} uses a method similar to the one used in \citet{krumholz}, but studies a region similar in scale to that of  \citet{bate09r}.  They also form  few objects ($\sim 15$).  They model a larger region with a lower initial density.  Their simulation differs from the work of  \citet{bate09r} because they include nuclear burning and accretion luminosity (both effectively ignored by \citet{bate09r}) which they state are the main sources of heating in their simulation.  Therefore, they agree that the work of \citet{bate09r} only shows the minimum possible effect of radiation on the star formation process.  
 
In our work, we attempt to model the effect of the heating of dust and gas by young stars on the form of the IMF in a clustered environment.  
We study a cluster forming region with scales an order of magnitude larger than previously studied by \citet{krumholz} ($\sim$ 100 AU to 1 pc). 
We are interested in the effect of our more realistic treatment of the temperature evolution on the mass function.  
Are previous works which ignore the local effect of forming stars realistic?  
Or is their work only applicable to the very earliest stages of star formation when starless cores are forming?  
In order to answer these questions, we simulate the evolution of a star-forming region with a hydrodynamics + gravity code and allow the sources which form within our region to ``turn-on'' and heat the surrounding material. 

The form of radiative transfer that we use is described in \citet{urban}.  
It differs from the form used by \citet{krumholz} and \citet{bate09r} in that our method does not assume gray radiation, is applicable to a range of optical depths, and assumes that the matter distribution around each source is spherically symmetric.  This latter approximation enables us to study a larger parameter space, which is necessary for modeling a clustered environment which is larger and where densities and optical depths are lower.  The method we use calculates the dust temperature from the source luminosity and the density distribution around the source, using a grid of models generated by the spherical radiative transport code DUSTY \citep{dusty}.

As in \citet{krumholz}, we assume that the gas and dust are effectively coupled and that dust heating controls the temperature in the region.  By ignoring other heating and cooling processes (such as cosmic rays and molecular cooling), we can examine in detail the effect of adding local heating that can not be described using an analytic approximation.  Our simulation with only dust heating provides a standard for future work.  We expect that including other cooling and heating mechanisms will only decrease the effect on the IMF that we will see in our simulation.  Therefore the results we show in this paper represent the case of dust heating as the dominant term in the energy equation at all densities and temperatures.

The simulations we present here are based on the work by MES06.  We run a total of four new simulations. For all of our simulations, we assume that the gas is isothermal until cores form at high densities.    Then we consider two different energy transfer methods.  In two of our simulations we assume that the gas is isothermal for the entire simulation runtime. In two different simulations, when the gas is dense and the first sink particle forms, we discard the assumption of isothermality and use the dust temperature calculation described in \citet{urban} to calculate the dust temperature and then set the gas temperature equal to the dust temperature.  For each isothermal simulation and each simulation with dust heating, we have two size/mass scales, small and large, determined by the number of generations of particle splitting.  We refer to these small and large simulations by $N_{\rm gen}=1$ or 2, respectively.  

In \S \ref{sec:code}, we discuss the algorithms we use to solve the fluid equations and  calculate the density profile, luminosity, mass accretion rate, and dust temperature.  In \S \ref{sec:sim}, we describe the details of the simulations.  In \S \ref{sec:results}, we describe our results.  This is followed by a discussion in \S \ref{sec:discuss} and our conclusions in \S \ref{sec:con}.

\begin{deluxetable}{crcr}
\tablecaption{Simulation Parameters}
\tablewidth{0pt}
\tablehead{
\colhead{$N_{\rm gen}$} & \colhead{$M_{\rm tot}[{\rm M}_\odot]$} &
\colhead{$L_{\rm box}[{\rm pc}]$} &
\colhead{$N_J$}
}
\startdata
 0 &  10.49 & 0.246 &   17 \cr
 1 &  83.92 & 0.492 &  136 \cr
 2 & 671.40 & 0.984 & 1088 \cr
\enddata
\label{tbl:params}
\end{deluxetable}

\section{The Numerical Algorithm}\label{sec:code}

We use the smoothed particle hydrodynamics (SPH) algorithm with particle splitting and sink particles that is described in MES06.  We have modified this algorithm to include the effect of the luminosity from forming stars on the dust/gas temperature in our simulation (see \S \ref{sec:lum} and \S \ref{sec:tdust}).

Other differences between this work and MES06 are the initial temperatures, 5K vs.\ 10K, and the threshold density contrasts for sink formation, 5,942 vs. 40,000, for this work and MES06, respectively.  
In the case of the isothermal simulations presented in MES06, the system modeled was scale-free.  
The parameters which determined the properties of the 
simulation were the temperature and density contrast.  
Choosing an initial density to interpret the results 
then set the scale of the simulation.  
By this method, the authors were able to scale their simulation to higher densities.  In our work, we fix the initial conditions of temperature and density for all simulations.   
Therefore, the inclusion of particle splitting, used to increase the resolution of the simulations in MES06, will instead be used to study larger, more massive regions.  
This is because our simulation does not increase the density  resolution as in the case of the isothermal simulation in MES06, 
but instead increases the size of our simulation box while the density resolution is held fixed for all simulations.   For our small simulation ($N_{\rm gen}=1$), we simulate a region of volume $\sim 0.5 $pc$^3$ with a mass of $\sim84 \msun$.  For our large simulation,  ($N_{\rm gen}=2$), the simulated region has a volume of $\sim1 $pc$^3$ and a mass of $\sim 670 \msun$.
Table \ref{tbl:params} illustrates the difference in size and mass of our simulations with increasing levels of particle splitting. 

\subsection{SPH}\label{sec:SPH}

We use a standard SPH algorithm (e.g., \citealt{monaghan}, and references therein), to simulate the evolution of a molecular cloud inside a cubic volume with periodic boundary conditions.  This algorithm was modified to include particle splitting and sink particles. The Jeans criterion requires that a Jeans mass contains a certain minimum number of SPH particles to be properly resolved, and prevent a spurious numerical effect known as artificial fragmentation (\citealt{truelove}, \citealt{boss}; see however \citealt{hubber}).  This translates into a condition between the mass, density, and specific internal energy of particles.  
Whenever a particle violates this condition, the code splits the particle, replacing it by 8 equal-mass particles located at the vertices of a cube (see \citealt{kw}, MES06).  Split particles can later re-split if the condition is violated again. 
We allow for a maximum of $N_{\rm gen}$ generations of splitting.  Hence, the mass ratio between the most and least massive particles is $8^{N_{\rm gen}}$.  

Sink particles (or sinks) are created when the gas density exceeds a density threshold $\rho_c$. A group of particles, whose total mass equals one Jeans mass, are replaced by a single, massive sink particle, which has the ability to grow by accreting gas particles.  
Boundary conditions are imposed at the interface between the sink and the surrounding gas.  In MES06, we used the boundary conditions described in \citet{bbp}. For this paper, we switched to the boundary conditions described in \citet{bcl}. 
We should point out that the particular choice of boundary conditions is not critical in our simulations, because infall of gas onto sinks tends to be radial and supersonic, making boundary conditions irrelevant \citep{bbp}. It was easier to implement the  dust physics (described in \S \ref{sec:tdust}) into the algorithm if we use the boundary conditions of \citet{bcl}, which is why we made the switch.  
For details, we refer the reader to MES06 and \citet{bcl}.

The sink particles in our simulation represent a star-forming core which may form a single star or a group of stars.  Fragmentation within a sink particle due to processes that may occur at higher temperatures and densities are not modeled in our simulation.  Therefore the masses of the sinks that we discuss throughout this paper only directly correspond to individual stars if no further fragmentation occurs within a sink.

\subsection{Density Profile} \label{sec:dens}

In order to determine the dust temperature in our simulation (as described later in \S \ref{sec:tdust}), we must first calculate the density profile by a spherical average around individual sinks.  To do this, we bin the particles around each sink into concentric shells which hold exactly 25 gas particles each, and calculate the density within each shell.  

The outer edge of the density distribution is set such that the mean density inside the outer radius is 200 times the initial background density.  (This method is commonly used in similar cosmological simulations, where this outer radius is referred to as the \textit{virial radius}; see,  e.g., \citealt{nfw}.)  We assume that it takes a minimum of 200 gas particles inside the outer radius to accurately determine the density profile.  If this condition is not met, then we cannot calculate the density profile (we discuss the effect of this in \S \ref{sec:tdust}).  We show examples of the density profile evolution in \S \ref{sec:ddpe}.  The density profile is parameterized by $n_o$ and $\alpha$,
\begin{equation}
n = n_o \left(\frac{r}{1000 \textrm{AU}} \right)^{-\alpha} \textrm{cm}^{-3}.
\end{equation}
Throughout this paper we will use the term, $n$,  to represent the number density of all particles ($n = n_{\rm{H_2}} + n_{\rm{He}}$) assuming $n_{\rm{H}_2}/n_{\rm{He}}=5$, which gives $\mu = 2.33$ (this relates the number and volume density discussed in \S \ref{sec:IC}).

\subsection{Luminosity and Mass Accretion Rate}\label{sec:lum}

In order to calculate the luminosity of a sink particle, we assume that it represents a single star with the mass of the sink particle.  We determine the luminosity of sink particles using the calculations of \citet{wt}, specifically their Table 3.  
\citet{wt} model the very early stages of stellar evolution.  In order to calculate the luminosity of their stars, they include accretion and contraction luminosity as well as stellar evolution models to calculate the luminosity from deuterium burning.  To calculate the accretion luminosity, they compute the accretion rate from fluid equations.  This differs from other pre-main sequence models which either ignore the mass accretion rate (\citealt{dm}; \citealt{siess})  or fix the mass accretion rate in their models \citep{zy}.  
\citet{wt} define a mass of the optically thick region in their models,  $M_\tau$.  We assume that our sink particle mass is equivalent to their $M_\tau$.  
We assume that the value of $M_\tau$, defined as the mass of the optically thick region by \citet{wt}, is equivalent to our sink particle mass. Therefore, given an $M_\tau$ and mass accretion rate, we can use their models to determine the time-averaged luminosity due to accretion and deuterium burning.
 
In this work, we assume that objects with $M\le 0.01 \msun$ do not have a high enough luminosity to affect the temperature of the surrounding gas; therefore, for these objects we set $L=0$.  For sink masses of $0.01<M\le0.04 \msun$, we assume that the luminosity is only a function of the sink mass and is independent of the mass accretion rate, $\dot{M}$.  We assume the luminosity follows the form 
\begin{equation}
\log (L/L_{\odot})=3.5 \log(M/\msun)+5.
\end{equation}
This relation can be derived from Figure 4 of \citet{wt}.  For objects with $M>0.04 \msun$, the luminosity is a function of $M$ and $\dot{M}$.  We determine $\dot{M}$ at a given time by calculating the total mass accreted by the source over the previous 5,000 years.

For objects with masses greater than 2$\msun$, we need to calculate the luminosity even though it is not given in the work of \citet{wt}.  We calculate the luminosity by assuming that the star is contracting on a Kelvin-Helmholtz timescale.  The luminosity of a young star depends on the combination of luminosity from nuclear burning, contraction, and accretion.  Models of pre-main sequence evolution of intermediate- and high-mass stars (e.g., \citealt{ps}) show that, without accretion, young massive stars evolve at constant luminosity before they reach the main sequence.  Therefore we assume that the luminosity (ignoring accretion) of a massive star is equal to its main sequence luminosity.  We define the total luminosity of a star with $M>2\msun$ as a combination of its main sequence luminosity and its accretion luminosity, $L=L_{\rm M.S.}+L_{\rm acc}$

To calculate the main sequence luminosity we use the relation $L/\lsun \sim (M/\msun)^{3.7}$ from \citet{zy}.  We summarize our calculation in Table \ref{tbl:lcalc}.  
In order to calculate the accretion luminosity we use the properties of the star from the previous timestep.  
We represent these values as primed symbols, i.e., $L'$, $L'_{\rm acc}$, $L'_{\rm M.S.}$, $M'$, $\dot{M'}$, $R'_{\rm acc}$, and $R'_{\rm M.S.}$, total luminosity, accretion luminosity, 
main sequence luminosity, mass, mass accretion rate, accretion radius (the actual radius of the object), and main sequence radius 
(the radius of the star assuming it is not accreting and has reached the main sequence), respectively, all in solar units.  
At the timestep when $M$ becomes greater than $2\msun$, the known values are $L'$, $M'$, and $\dot{M'}$.  
The values that we derive are $L'_{\rm acc}$, $L'_{\rm M.S.}$, $R'_{\rm acc}$, and $R'_{\rm M.S.}$.  The main sequence luminosity is $L'_{\rm M.S.}/\lsun = (M'/\msun)^{3.7}$ 
and the accretion luminosity is $L'_{\rm acc} = L' - L'_{\rm M.S.}$.
The main sequence radius can be calculated using $R'_{\rm M.S.}/R_{\odot}=(M'/\msun)^{0.5}$ (from \citealt{zy}).  The accretion radius is $R'_{\rm acc}=G M'\dot{M'} /L'_{\rm acc}$.    
These values are used to calculate the contraction timescale later.

\begin{deluxetable}{lllll}
\tablecaption{Luminosity Calculation}
\tablewidth{0pt}
\tablehead{
\colhead{Known} &\colhead{}&
\colhead{Derived}
}
\startdata
$L'$, $M'$, $\dot{M'}$ & &$L'_{\rm M.S.}$ &=& $(M')^{3.7}$\cr
&& $L'_{\rm acc}$& =& $L' - L'_{\rm M.S.}$ \cr
&& $R'_{\rm M.S.}$ & =& $(M')^{0.5}$\cr
&& $R'_{\rm acc}$ & =& $G M' \dot{M'} / L'_{\rm acc}$ \cr
\hline
&$\Delta t$& \cr
\hline
&& $L_{\rm M.S.}$ & =& $M^{3.7}$\cr
$M$, $\dot{M}$  & & $t_{\rm KH}$ & =& $3M'^2 G/5 R'_{\rm acc} L$'\cr
&& $dR/dt$ & =& $(R'_{\rm acc} - R'_{\rm M.S.})/t_{\rm KH}$\cr
&& $R_{\rm acc}$ & =& $R'_{\rm acc} - dR/dt \times  \Delta t$\cr
&& $L_{\rm acc}$ & =& $GM\dot{M} /R_{\rm acc}$\cr
&& $L$& =& $L_{\rm M.S.} + L_{\rm acc}$
\enddata
\label{tbl:lcalc}
\end{deluxetable}

We now have all of the parameters needed to calculate $L$ at the current timestep (un-primed symbols represent values at the current timestep). Therefore, $L=L_{\rm M.S.} + L_{\rm acc}$.  $L_{\rm M.S.}$ can be calculated from the mass.  However, $L_{\rm acc}$ is more difficult to calculate.  We attempt to include the contraction luminosity through the contraction timescale when we calculate the accretion luminosity; for a known main sequence and accretion luminosity, we assume that a star contracts on the Kelvin-Helmholtz timescale, 
\begin{equation}
t_{\rm KH}=\frac{3M'^2 G}{5R'_{\rm acc}L'}.
\end{equation}
Then we can calculate the rate of change from the previous accretion radius to the main sequence radius, $dR/dt = (R'_{\rm acc}-R'_{\rm M.S.})/t_{\rm KH}$.  Now that we have the rate of change of the radius, we can calculate the new accretion radius, $R_{\rm acc}=R'_{\rm acc}- dR/dt \times \Delta t$, where $\Delta t$ is the change in time between the previous and current timestep in the simulation.  Then we can derive the new accretion luminosity, $L_{\rm acc}=GM\dot{M}/R_{\rm acc}$. Combining the new accretion luminosity and new main sequence luminosity gives us the new total luminosity.  As we discuss later in \S \ref{sec:mlevol}, we found that this approach leads to a smooth transition from $M < 2\msun$ to $M > 2\msun$, with no discontinuities in the luminosity.
 
 \subsection {Dust and Gas Temperature}\label{sec:tdust}

The temperature of each SPH particle is initially set to 5K.  When the first sink particle forms and has a non-zero luminosity, then we begin calculating the dust temperature.  In order to calculate the dust temperature in our simulation we use the calculation described in \citet{urban}.  This method uses a spherical continuum radiative transfer code to calculate the dust temperature around individual sources with a given luminosity and surrounding, envelope density profile.  Using the temperature profiles around individual stars to calculate the flux at every point in the simulation, the dust temperature can be calculated.  The gas temperature is then set equal to the dust temperature.

Some differences between the method described in \citet{urban} and the method we use in this paper are discussed.  Instead of a polynomial interpolation method, we use weighted-linear interpolation to decrease the computational time spent on this step.  We conducted tests similar to those performed in \citet{urban} to test our new interpolation method.  We do not find significant differences between the two interpolation methods.  

We assume that the outer radius of our dust-gas envelope is 0.1pc for all of the sink density profiles.  This is because the density profile derived using the method described in \S \ref{sec:dens} always derived a outer radius less than 0.1 pc.  We found in \citet{urban} that the outer radius does not have a significant effect on the dust temperature.  Therefore, we are not concerned that this assumption will strongly affect our dust temperature calculation.

We have also needed to extrapolate in a few cases when the parameters of our sink particles were outside of the parameter space studied in \citet{urban}.   
In  \S \ref{sec:ddpe} and \S\ref{sec:mlevol}, we discuss the sink properties that determine the dust temperature - the envelope's density profile and the sink particle's luminosity.  
Based on the values of luminosity (i.e. $L<10^6\lsun$), it is unlikely that extrapolation occurred due to luminosities that were outside the parameter space.  
However, this is not the case for the density profile.  For $\alpha>2.0$ and log $n_o>6$, extrapolation was more frequent.   
But this extrapolation only occurred in the range of $\alpha$ between 2 and 2.5 and for values of log $n_o $  between 6 and 8.  
Figures 12 and 13 from \citet{urban} show that for this range of $\alpha$ and log $n_o$ the behavior of $K$ and $\beta$ (the parameters that determine the dust temperature) is somewhat regular for low luminosities.  
For higher luminosities ($L>10^4 \lsun$), the behavior is not as regular.  This may introduce some uncertainty in our temperatures around high luminosity sources.

In some cases we were unable to derive a density profile because the sink particle was surrounded by too few gas particles.  When this occurred, we assumed $\alpha = 0$ and did not calculate a density profile for that sink.  Therefore, at that timestep, the sinks with no density profiles were not used to calculate the dust temperature in the simulation.   In some cases, the value of $\alpha$ was found to be less than 0.5.  When this occurred, we simply assumed that the value of $\alpha$ was 0.5 and used that value to calculate the value of $n_o$.  This assumption must be made and is due to our general assumption of spherically distributed material around sink particles.  This approximation does not occur very often in our simulation as can be seen later in Figures \ref{fig:anoevol1} and \ref{fig:anoevol2}.

\section{The Simulations}\label{sec:sim}

We perform four different simulations.  They  incorporate all of the same physical processes (i.e., hydrodynamics and gravity) and are identical in all aspects excepting the following two. (1) Two of the simulations assume an isothermal equation of state at a fixed temperature of 5K throughout the entire calculation ($T_{\rm gas} = 5$K).  The other two simulations include the effect of dust heating discussed in \S \ref{sec:tdust} ($T_{\rm gas} = T_{\rm dust}$). 
For each method of calculating the temperature, we simulate a small and a large region by changing the level of particle splitting.
 (2) Two simulations allow one generation of particle splitting (N$_{\rm gen}=1$) and two simulations allow two generations of particle splitting (N$_{\rm gen}=2$).  The size and mass of these simulations depend on the number of generations of particles splitting and the values are presented in Table \ref{tbl:params}.  (The values for a simulation with no particle splitting, N$_{\rm gen}=0$, are shown only for reference.)  A higher level of particle splitting allows larger and more massive regions to be simulated. 

Our SPH algorithm simulates the evolution of a region deep within a molecular cloud as a cubic volume  with periodic boundary conditions, containing initially $64^3$, or $262,144$ particles. 
The volume of the box is $L_{\rm box}^3$; see Table \ref{tbl:params} and \S\ref{sec:IC} for details on $L_{\rm box}$.
Particle splitting, sink formation, and accretion onto sinks are included. For details of the SPH algorithm, we refer the reader to MES06 and \S \ref{sec:code}.

\subsection{Initial Conditions and Simulation Parameters}\label{sec:IC}

Our method for generating initial conditions is described in detail in MES06.  We start with a uniform density distribution, with no overall density gradient.  Onto this we superpose a small density perturbation which is described by a Gaussian random field with a density power spectrum $P(k)\propto k^{-2}$, where $k$ is the wavenumber.  In terms of the simulation setup, these initial conditions are achieved by arranging the SPH particles on a $64\times64\times64$ cubic grid, and slightly displacing them to reproduce the desired power spectrum.

To generate initial conditions, we first need to fix the mass resolution of the algorithm. In their SPH simulations, \citet{bb05} use a barotropic equation of state, which is isothermal at densities $\rho\leq10^{-13}{\rm g\,cm^{-3}}$ (or $n\approx 2.6\ee{10}$cm$^{-3}$), and adiabatic at densities $\rho>10^{-13}{\rm g\,cm^{-3}}$. 
In this case, there is a minimum Jeans mass $(M_J)_{\min}=0.0011\,{\rm M_\odot}$, corresponding to the density $\rho=10^{-13}{\rm g\,cm^{-3}}$.  Since we focus on the formation of a star cluster and not the details of individual star formation, it is unnecessary for us to resolve such high densities and low masses. Instead, we set the mass resolution limit of our algorithm at $M=0.008\,{\rm M_\odot} \approx 10\% $ of the minimum mass for hydrogen-burning.  
This is not much of a limitation, since the effect of dust becomes important only at much larger masses/higher luminosities.  
A drawback of this limited resolution is that we will not be able to assess the effect of heating by massive stars on the subsequent formation of objects with masses below $0.008\,{\rm M_\odot}$.  
However, this lower resolution allows us to simulate a large region which contains enough mass to potentially form several high-mass stars.

We consider a cloud with initial density $\bar\rho=4.75\times10^{-20}{\rm g\,cm^{-3}}$, 
or $\bar n=1.22\times10^4{\rm cm}^{-3}$ 
assuming $\mu = 2.33 $ ($\rho = \mu n m_{\rm H}$), and temperature $T=5$K (see Table \ref{tbl:dens}).  Our choice of an initial temperature of 5K, as opposed to 10K used in other simulations (\citealt{bbb}; MES06), was motivated by the low temperatures in the calculations of \citet{urban} as well as the discussion in \citet{larson} and recent observations from \citet{evans} and \citet{crapsi}. 
The gas will remain isothermal at 5K, until it is heated by dust.

\begin{deluxetable*}{lcc}
\tablecaption{Density Parameters}
\tablewidth{0pt}
\tablehead{
\colhead{Density} & \colhead{g cm$^{-3}$}  &\colhead{cm$^{-3}$}
}
\startdata
Initial average density ($\rho_i$) &  $4.75\ee{-20}$ &   $1.22\ee{4}$ \cr
Sink creation density  &  $2.82\ee{-16}$ & $7.25\ee{7}$  \cr
\hline
\hline
Splitting Density for N$_{\rm gen} =1$ & $371 \times \rho_i$ &  \cr
1st and 2nd Splitting Density for N$_{\rm gen} =2$ &  $5.8 \times \rho_i$& $371 \times \rho_i$ 
\enddata
\label{tbl:dens}
\end{deluxetable*}

Our algorithm will turn dense gas clumps of mass $M=0.008\,{\rm M_\odot}$ into sinks. To justify this, these objects must have a mass equal to the Jeans mass. For a gas with polytropic constant $\gamma=5/3$, the Jeans mass is given by 
\begin{equation}\label{eq:mj}
M_J=\biggl({5kT\over2G\mu}\biggr)^{3/2}
\biggl({4\pi\rho\over3}\biggr)^{-1/2}\,
\end{equation}
\citep{tohline}.
Since the gas is isothermal, we can set $T=5\,{\rm K}$ and $M_J=0.008\,{\rm M_\odot}$ in equation~(\ref{eq:mj}), and solve for the density. 
We get $\rho_c=2.822\times10^{-16}$g cm$^{-3}$, or $n_c=7.252\times10^{7}$cm$^{-3}$.  This is the threshold density at which sinks will be created (see Table \ref{tbl:dens}).  
This will happen after the gas contracts from the initial density, $\bar\rho$, by a factor of $\rho_c/\bar\rho=5,942$. 
This factor is smaller than the value of 40,000 used in MES06, but still provides a wide dynamical range in density.

It takes a minimum number of particles to properly resolve a Jeans mass. The precise value depends on the particular implementation of SPH. In MES06, we assumed a typical value of 100 particles.  In this paper, we also split particles when the Jeans mass drops below 100 particles, but we require that the clumps that turn into sinks contain 200 particles. The reason is that clumps with 100 particles often fail the criteria for sink creation because their rotation or internal motions make them unbound.
Therefore, the mass of an SPH particle is 
\begin{equation}
m_{\rm part}={0.008\,{\rm M_\odot}\over200}=4\times10^{-5}{\rm M_\odot}\,.
\label{masspart}
\end{equation}
Our simulations start with $64^3$ particles.
Therefore, the total mass of the system is $M_{\rm tot}=64^3m_{\rm part}=10.49\,{\rm M_\odot}$, and the box size is $L_{\rm box}=(M_{\rm tot}/\bar\rho)^{1/3}=0.246\,{\rm pc}$.
By setting $\rho=\bar\rho$ and $T=5{\rm K}$ in equation (\ref{eq:mj}), we get an initial Jeans mass $M_{J,{\rm init}}=0.617\,{\rm M_\odot}$.  Hence the system starts up with $N_J=17$~Jeans masses, compared to 500 in MES06.  However, these numbers assume no particle splitting.   If we allow particles to split, each splitting generation will increase the effective number of particles by a factor of 8.  The densities at which particle splitting occurs are listed in Table \ref{tbl:dens}.   Since the mass of the final generation of particles is fixed by equation~(\ref{masspart}), each splitting generation will increase $M_{\rm tot}$ and $N_J$ by a factor of 8 and $L_{\rm box}$ by a factor of 2.  With a fixed density resolution and a fixed minimum mass resolution, we effectively increase the size and mass of our simulated region with each new level of particle splitting.  To get systems with reasonable sizes and masses, we allow for 1 and 2 generations of particle splitting, $N_{\rm gen} = 1$ and 2. 

As mentioned before, our density resolution is not as high as that of MES06.  If we compare to Case 4 listed in Table 4 of MES06, we find that they start with a higher initial density and model a smaller region (L$_{\rm box}=0.38$pc) with a larger mass (M$_{\rm total} = 320 \msun$) than our N$_{\rm gen}=1$ model.  However, our N$_{\rm gen}=2$ model is larger in volume and more massive, but still starts with a smaller initial density.

\subsection{Sink Particles}

As mentioned in the previous section, sink particles are formed from 200 gas particles which exceed the threshold density for sink creation ($\rho_c=2.822\ee{-16}$ g cm$^{-3}$).  The second criterion for sink creation is that the gas particles must also be Jeans unstable, meaning that they are collapsing and have formed a gravitationally bound system at the time that the sink is created.  The exact prescription for creating a sink particle is described below.

We define an {\it accretion radius} $r_{\rm acc}$, such that when a particle reaches the threshold density $\rho_c$, a sphere of radius $r_{\rm acc}$, centered on the new location of the sink particle, will contain 200 gravitationally bound particles.  In the isothermal simulations, this corresponds to a Jeans mass. These 200 particles are then removed and replaced by a sink particle with the same total mass and center-of-mass position and velocity as the 200 particles.  In order to determine the value of the accretion radius, we have run test cases in the isothermal simulation in which we vary the value of the accretion radius and allow a few sink particles to form.  We choose the value of the accretion radius that ensures that the number of particles that are used to create a sink particle is approximately 200.  For our simulations the accretion radius is set at $\sim 150$AU.

For the runs that include dust heating, the temperature will vary throughout the simulation so there will be no fixed value of the Jeans mass.  In this case, we still use the accretion radius from the isothermal simulation.   The first criterion of needing $\sim 200$ particles will still be met if we use the isothermal accretion radius.  
However, the second criterion requiring that the particles must be Jeans unstable before a sink forms will tend to delay sink formation compared to the isothermal simulation.  This is because the Jeans mass in a simulation with dust heating will typically  be higher due to the increased temperature.  Hence, it will typically take more than a collection of 200 Jeans-unstable gas particles to have a total mass equal to the Jeans mass;  
when a particle reaches the threshold density, sink formation will be delayed until enough particles (or mass) are within the accretion radius to increase the enclosed mass to the local Jeans mass defined by the local temperature. 

Once sink particles are formed, they have the ability to grow by accreting gas particles. Whenever a gas particle enters the accretion radius of a sink, that particle will be accreted, provided that it is gravitationally bound to the sink.   Typically the sink particles will have time to accrete mass before their luminosity is large enough to heat the surrounding dust and gas.

\subsection{Timescales}\label{sec:timescales}

Since our initial density is $\bar\rho=4.75\times10^{-20}{\rm g\,cm^{-3}}$, or  $n=1.22\times10^4{\rm cm}^{-3}$, the initial free-fall time, defined as 
\begin{equation}
\tff = \sqrt{3 \pi / 32 G \bar{\rho}},
\end{equation}  
is $\tff =  9.64\ee{12}$s $ = 3.06\ee{5}$yr. This is true for all simulations because the initial density is the same for all cases.

Since we do not include ionizing radiation in our simulations, they are no longer realistic when very massive stars form.  Therefore, we stop our simulations when the most massive sink particle in each reaches a mass of 20.8 \msun, which is approximately the mass of an O9.5 star.  \citet{keto} shows that for earlier spectral types, the HII region surrounding a star (or group of stars) could be greater than the size of our sink particle radius or would be expanding.  It is no longer realistic to ignore the effects of ionization.  Therefore we stop our simulation when we have reached this limit.  This occurs at 2.5$\tff$ for the N$_{\rm gen}=2$ simulation with dust heating.  We stop our N$_{\rm gen}=2$ isothermal simulation at the same time so that we can compare the two N$_{\rm gen}=2$  simulations.  
At $t=4.5 \tff$,  in our N$_{\rm gen}=1$ simulation, no sinks have yet reached 20.8 $\msun$.  However, over 80\% of the mass is in sinks.  Therefore, we stop the simulation at this point. 

\subsection{Gas Temperature}

Our version of SPH does not include the standard energy equation, with $p\,dV$ work, viscous heating, and radiative cooling. Instead, the temperature of each particle is set at each timestep, depending on the local conditions. In the isothermal runs, the temperature is kept fixed, while in the runs with dust heating, the gas temperature is set to the local temperature of the dust (see \S \ref{sec:tdust}). We run two isothermal simulations, and two simulations with dust heating, both with one and two generations of particle splitting, for a total of four simulations. In Table \ref{tbl:sim}, we list the four different simulations, and give the final number of sinks and the final mass in sinks.

\begin{deluxetable*}{c l c r c c c}
\tablecaption{Summary of Simulations}
\tablewidth{0pt}
\tablehead{
\colhead {$N_{\rm gen}$} & \colhead{$T_{\rm gas}$} &
\colhead{Final Time} & \colhead{Final \# of Sinks} 
& \colhead {Max. Sink Mass } 
& \colhead {Total mass in Sinks} &\colhead{SFR$_{\rm ff}$}\\
\colhead {} & \colhead{} &
\colhead{($\tff$)} & \colhead{} 
& \colhead {($\msun$)} 
& \colhead {\% }&\colhead{}
}
\startdata  
1  &5K                      & 4.5 & 518  & 0.46& 98\%  &  0.22\cr
1  &T$_{\rm dust}$ & 4.5 &   20  & 12.4 & 87\%  &  0.19 \cr
2  &5K                      & 2.5 &   3429&  0.50 & 60\%  & 0.24 \cr
2  &T$_{\rm dust}$ & 2.5 &  74   & 20.8 & 50\%  &  0.20\cr
\enddata
\label{tbl:sim}
\end{deluxetable*}

\section{Results}\label{sec:results}

\begin{figure}
\epsscale{1.0}
\plotone{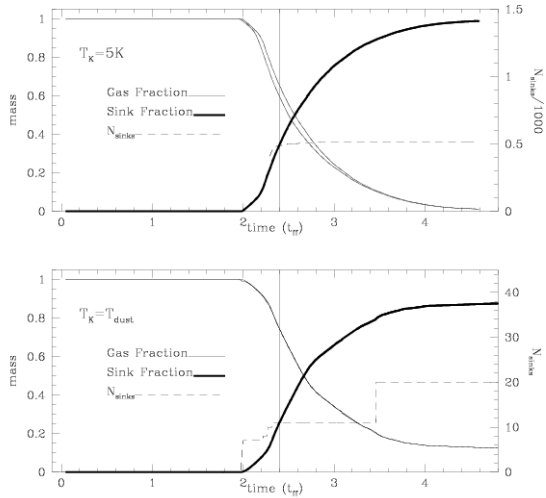}
\caption[Mass fraction evolution for simulations with $N_{\rm gen}=1$, Isothermal and Dust Heating. ]
{Mass fraction evolution for simulations with $N_{\rm gen}=1$.  Panels show fraction of mass in gas and sinks as a function of free-fall time.  Top panel shows results for the isothermal simulation.  Bottom panel shows results for simulation with dust heating.  Thick line shows the mass fraction in sinks.  Thin lines show the mass fraction in gas as a stacked histogram for different generations of particles.  Lower thin line represents particles which have not undergone particle-splitting.  Higher thin line represents fraction of particles which have split once.  The two gas lines are barely indistinguishable for the simulation with dust heating.  Dashed line shows the evolution of the number of sinks.  
Vertical line at 2.4$\tff$ is shown for reference.}
\label{fig:frac1}
\end{figure}

\begin{figure}
\epsscale{1.0}
\plotone{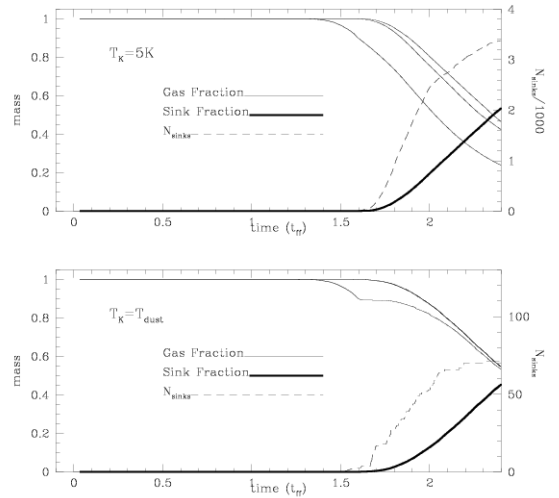}
\caption[Mass fraction evolution for simulations with $N_{\rm gen}=2$, Isothermal and Dust Heating. ]
{Mass fraction evolution for simulations with $N_{\rm gen}=2$.
See Figure \ref{fig:frac1} for details of plot. 
The line representing particles that have split twice is 
barely indistinguishable from the line representing particles 
that have split once in the bottom figure. }
\label{fig:frac2}
\end{figure}

Figures \ref{fig:frac1} and \ref{fig:frac2} show how the mass in the simulations is distributed between the gas and the sink particles as a function of time.  These figures show that as the mass in gas particles decreases, the mass in sink particles increases, as expected.  It is also interesting to note that the transition from a gas-dominated to sink-dominated simulation (point at which the sink and gas lines cross in Figs. \ref{fig:frac1} and \ref{fig:frac2}) occurs at different times for the different simulations .  
Since dust heating increases the average temperature of the simulation box (see \S \ref{sec:tempdens} below) and thus inhibits the formation of sink particles, 
the transition from gas-dominated to sink-dominated occurs at a later time in simulations with dust heating, both for N$_{\rm gen}=1$ and N$_{\rm gen}=2$.  Indeed, for the case N$_{\rm gen}=2$ with dust heating, the mass in sinks has not yet reached 50\% by the end of the simulation, at $t=2.5 \tff$.  

The increase in temperature that inhibits the formation of sink particles also affects particle splitting.  In the simulations with dust heating, the higher temperature leads to a higher local Jeans mass.  Therefore, the condition which triggers particle splitting, i.e. violating the Jeans criterion, is rarely met at the highest level of particle splitting for the simulations with dust heating, unlike the case for the isothermal simulations.  This can be seen in the lines representing ``Gas Fraction'' in  Figures \ref{fig:frac1} and \ref{fig:frac2}, as well as the lack of blue and green dots in the right-hand panels of Figures \ref{fig:xy1} and \ref{fig:xy2}, respectively.

Another interesting feature of Figures \ref{fig:frac1} and \ref{fig:frac2} is the large difference in the number of sink particles formed in the simulations.  This can also be seen in Tables \ref{tbl:sim1} and \ref{tbl:sim2}.  The isothermal simulations form more than an order of magnitude more sink particles than the equivalent simulations with dust heating.  However, as seen in Table \ref{tbl:sim}, the amount of material in sinks is comparable for simulations with similar sizes, i.e., same values of  N$_{\rm gen}$.  We discuss these two features next.

\begin{deluxetable*}{c|cc|cc}
\tablecaption{Simulation Summary: $N_{\rm gen}=1$ }
\tablewidth{0pt}
\tablehead{
\colhead{time } & \colhead{$T_{\rm gas} = 5$K} 
& \colhead{}& \colhead{$T_{\rm gas} = \td$} &\colhead{} \\
\colhead{($\tff$)} & \colhead{Max. Mass ($\msun$)} & \colhead{\# of Sinks}& \colhead{Max. Mass ($\msun$)} & \colhead{\# of Sinks}
}
\startdata  
2.0  & 0.03    &   17  & 0.031& 7 \cr       
2.5  & 0.23   &   500  &5.281&11\cr 
3.0  & 0.31   &   518  & 10.002& 11\cr
3.5  & 0.41   & 518   &11.907 &20\cr
4.0  &0.46   &   518   & 12.316& 20\cr
4.5  & 0.46   &  518   &12.375 &20\cr 
\enddata
\label{tbl:sim1}
\end{deluxetable*}

\begin{deluxetable*}{c|cc|cc}
\tablecaption{Simulation Summary: $N_{\rm gen}=2$  }
\tablewidth{0pt}
\tablehead{
\colhead{time } & \colhead{$T_{\rm gas} = 5$K} 
& \colhead{}& \colhead{$T_{\rm gas} = \td$}  & \colhead{}\\
\colhead{($\tff$)} & \colhead{Max. Mass ($\msun$)} & \colhead{\# of Sinks}& \colhead{Max. Mass ($\msun$)} & \colhead{\# of Sinks}
}
\startdata  
1.6   &0.04     &     14&   0.066  &  3   \cr       
1.8   &0.11    &     899&  2.11   & 28       \cr 
2.0   &0.20     &   2452& 7.99  & 53        \cr
2.2   &0.34   &     3033& 12.40  & 70     \cr
2.4   &0.46    &    3371&  17.99  & 71     \cr
2.5   &0.50    &  3429   &  20.8  &   74 \cr
\enddata
\label{tbl:sim2}
\end{deluxetable*}

The difference in number of sinks formed is strongly affected by dust heating.  As the sink particles heat the simulation volume in the simulation with dust heating, the growth of structure and sink particle formation is inhibited.  The heated gas prevents the fragmentation which occurs unhindered in the isothermal simulation.  Since the percentage of the total mass in sinks is comparable for simulations with the same value of N$_{\rm gen}$ and there are fewer sinks in the simulations with dust heating, the sinks in the simulations with dust heating are on average much more massive than the sinks in the isothermal simulations.  Gas that is prevented from forming new sinks instead accretes onto existing ones. This will be discussed in \S \ref{sec:mlevol}  below.

The percentage of material in sink particles at the end of the simulations, or the star formation efficiency, is affected slightly by the energetics algorithm used, i.e. isothermal or with dust heating.  
As seen in Table \ref{tbl:sim}, in the N$_{\rm gen} = 2$ calculation, 60\% of the mass is in sinks after 2.5 free-fall times for the isothermal
calculation versus 50\% for the simulation with dust heating.
Another measure of the star formation efficiency (or the speed of star formation) has been defined by \citet{KM05}.  
SFR$_{\rm ff}$ is the fraction of material converted into stars per free-fall 
time \citep{KT07}.
\citet{KT07} argue that SFR$_{\rm ff}$  ranges over
0.013 - 0.03 for the Milky Way Galaxy based on the interpretation of
various observations, which is in agreement with the work of \citet{KM05}.  Studies of nearby clouds forming low-mass stars found an average value of 0.04 \citep{evans09}. 

We give the values of SFR$_{\rm ff}$ for our simulations in Table \ref{tbl:sim}.  
All of our results are about an order of magnitude higher than the prediction of \citet{KT07} and a factor of 5 higher than the results of \citet{evans09}.  
The values of SFR$_{\rm ff}$ for the simulations with dust heating are slightly lower than the values for the isothermal simulations.  
We believe the higher values of SFR$_{\rm ff}$ in our simulations probably arise because we are not including turbulence.
In the theory of \citet{KM05}, the main regulating agent of star formation is turbulence.  
Since we do not include this physical effect, which would slow star formation in our simulation, it is not surprising that we find higher values of SFR$_{\rm ff}$.

\subsection{Sink Particle and Gas Mass Distribution}

\begin{figure*}
\epsscale {1.0}
\plottwo{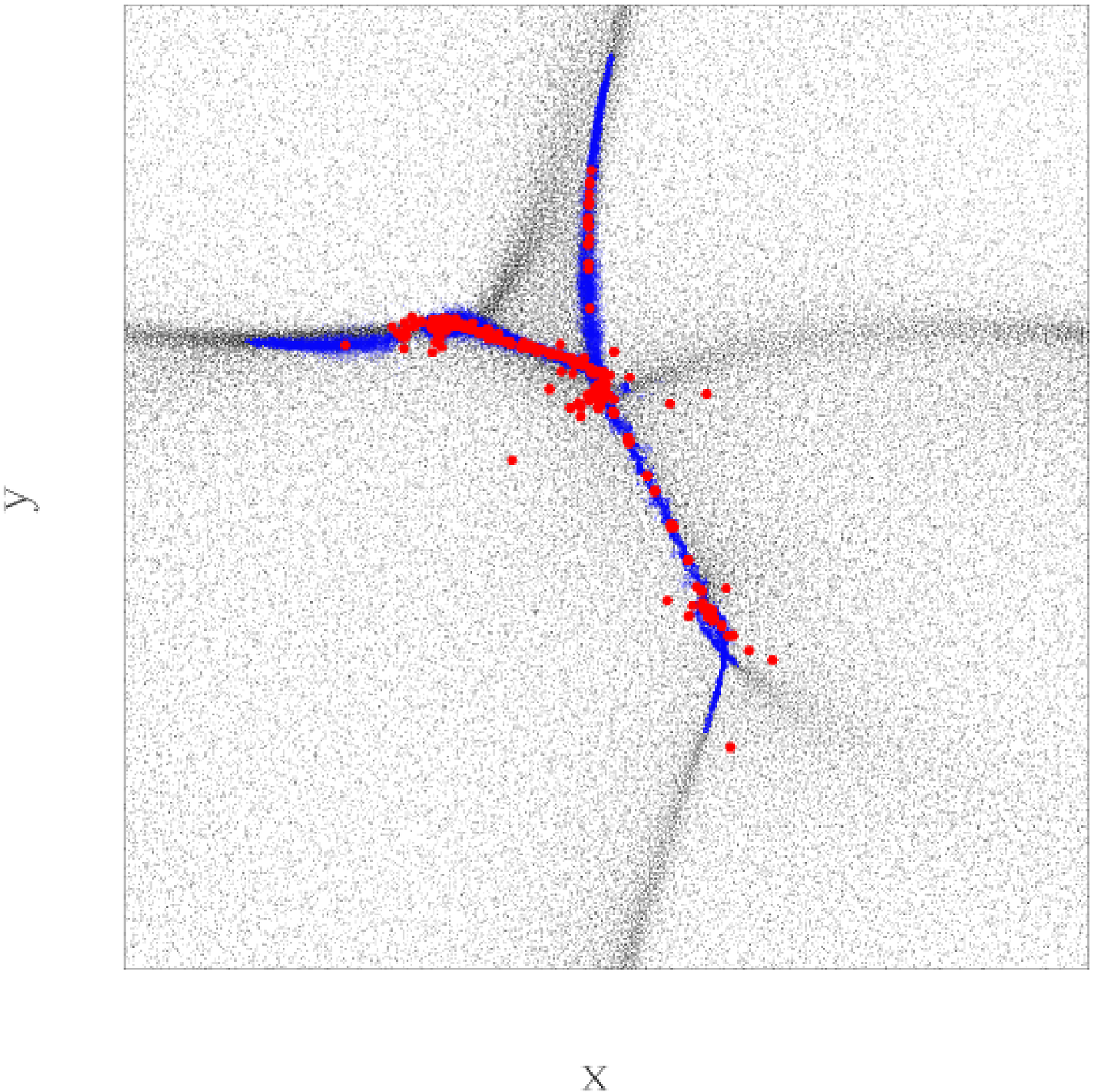}{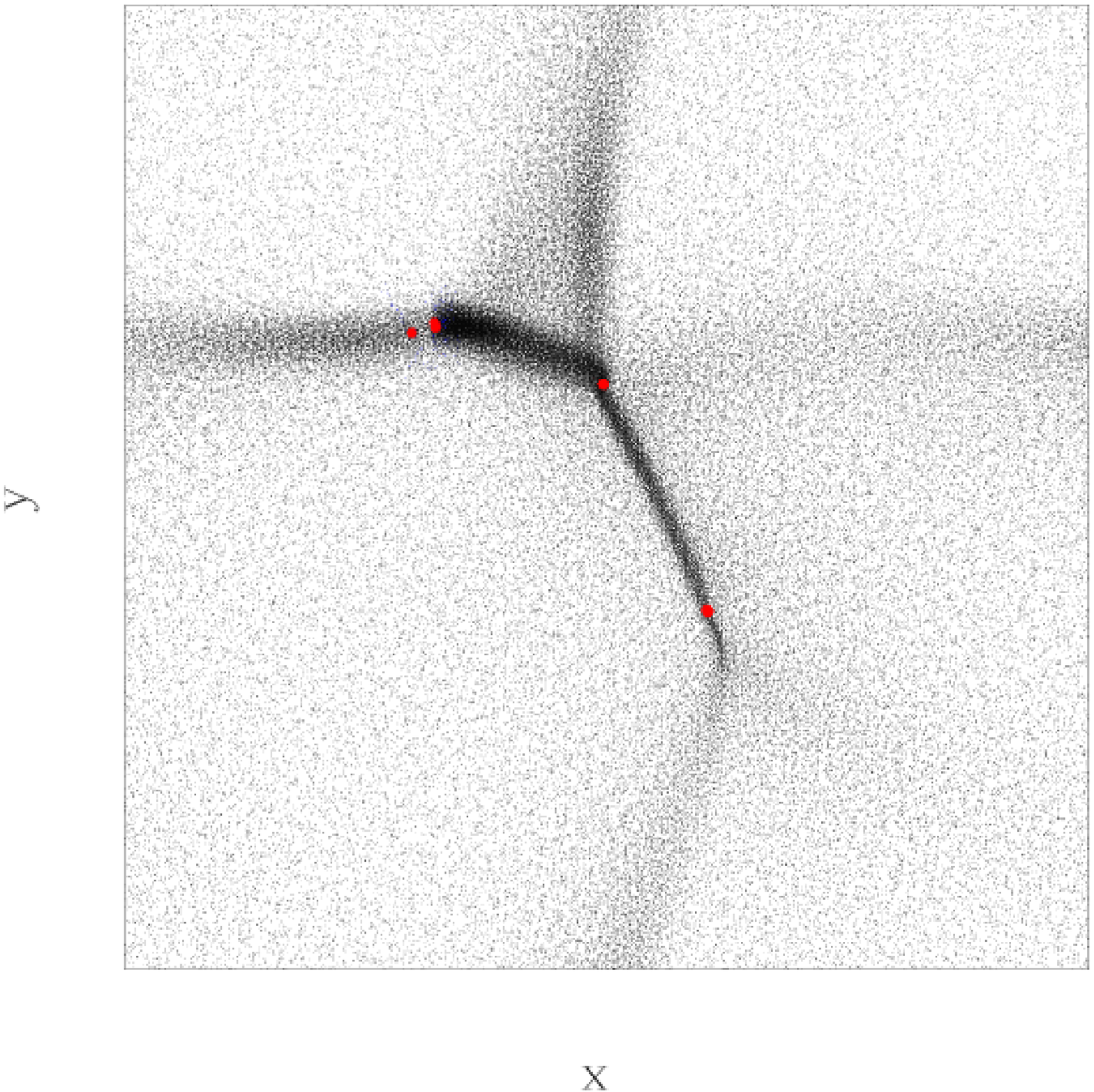}
\caption{XY position plot of sink and gas particles for simulations with N$_{\rm gen}=1$ at 2.4 $\tff$.  Left plot shows result from isothermal simulation.  Right plot includes dust heating.  Black and blue dots indicate gas particles.  Blue dots are gas particles which have undergone one particle splitting.  Red dots are sinks.  Scale is 0.492 pc. $\times$ 0.492 pc.}
\label{fig:xy1}	
\end{figure*}

\begin{figure*}
\epsscale {1.0}
\plottwo{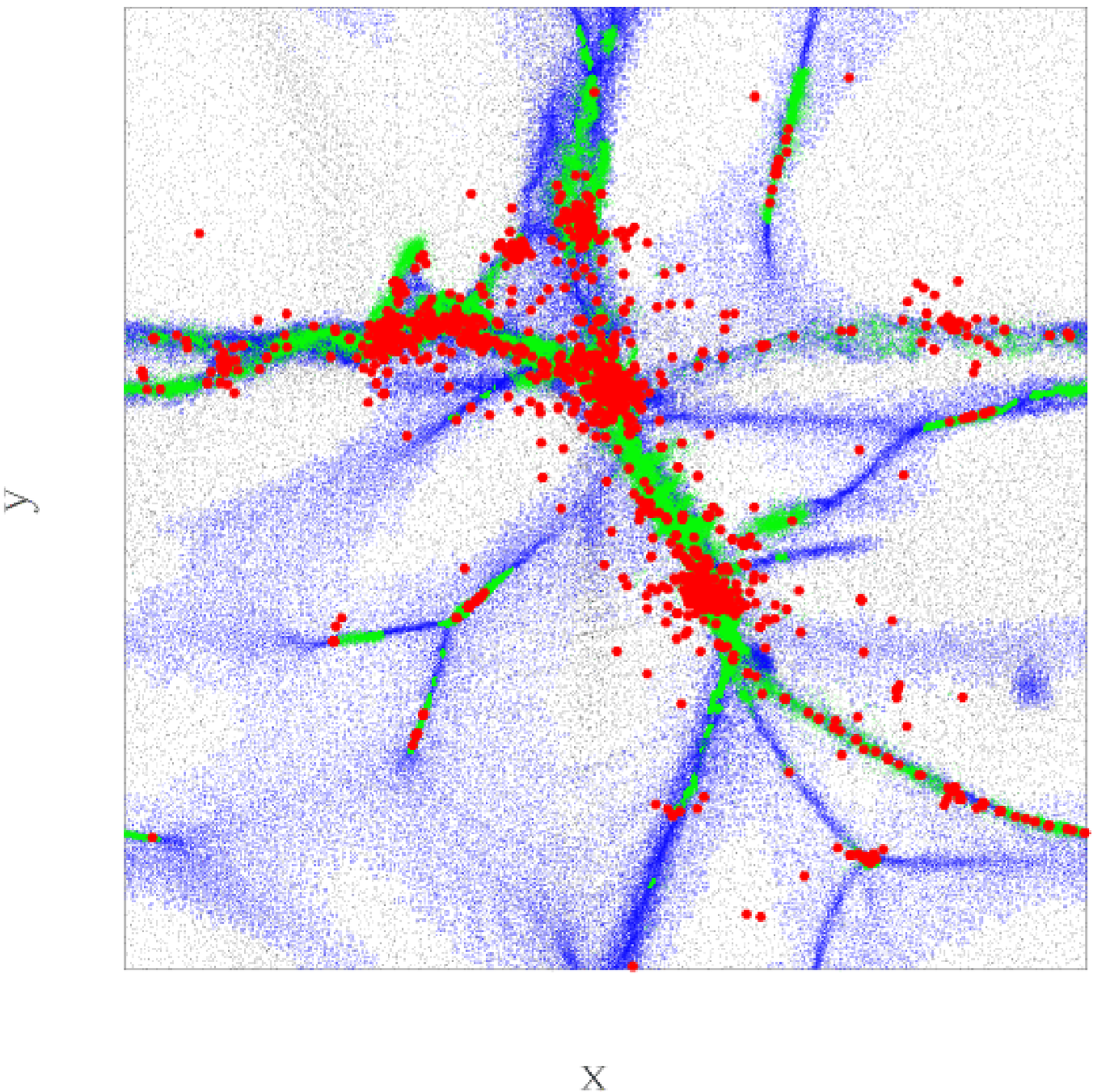}{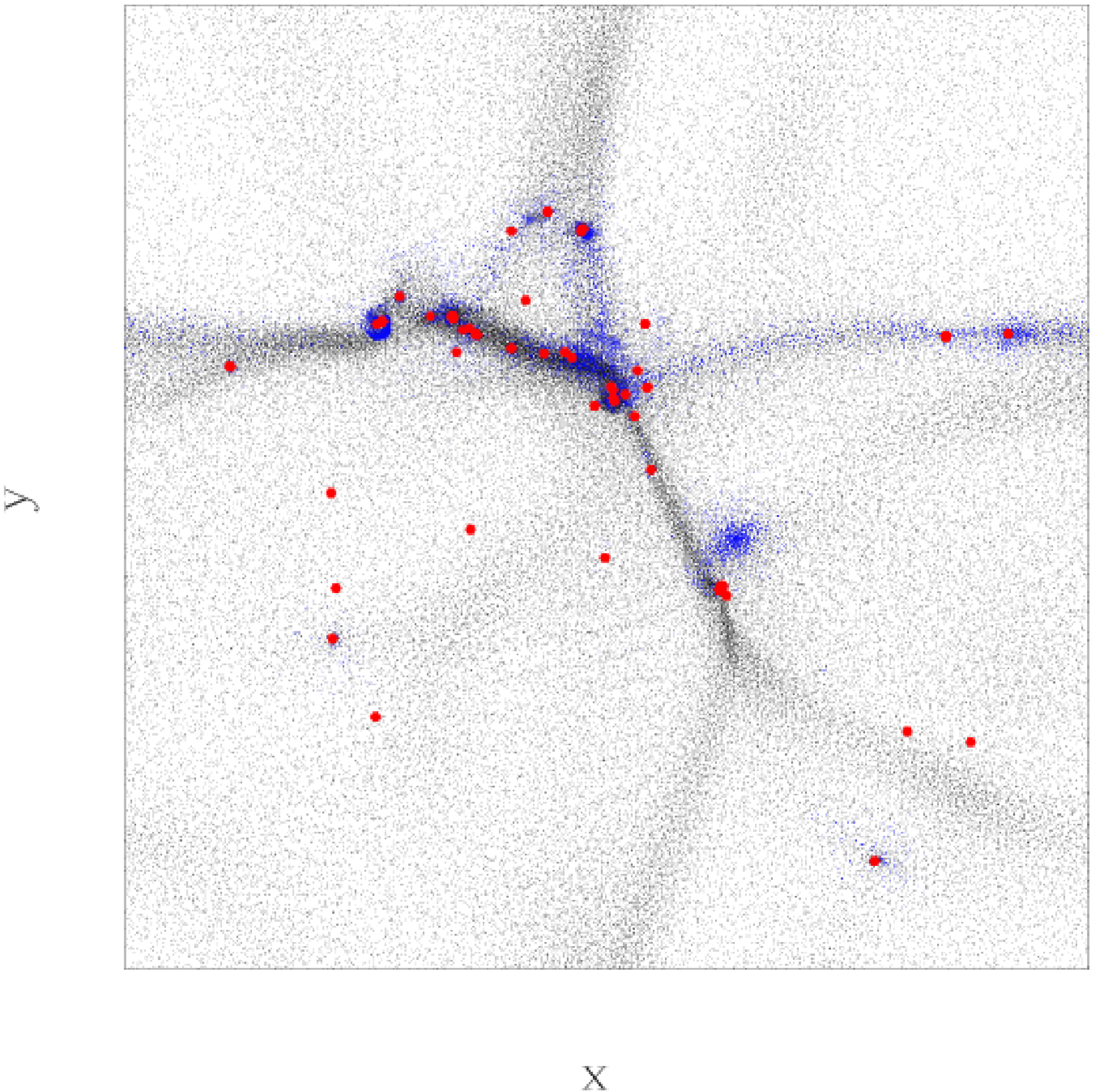}
\caption{XY position plot of sink and gas particles for for simulations with N$_{\rm gen}=2$ at 2.4 $\tff$.  Left plot shows result from isothermal simulation.  Right plot includes dust heating.  Black, blue, and green dots indicate gas particles.  Blue dots are gas particles which have undergone one particle splitting.  Green dots are particles which have split twice.  Red dots are sinks.  Scale is 0.984 pc $\times$ 0.984 pc.}
\label{fig:xy2}	
\end{figure*}

Figures \ref{fig:xy1} and \ref{fig:xy2} show the distribution of sink and gas particles in our simulation box.  They are shown at t =  2.4 $\tff$ because at this time a substantial fraction of the total number of final sink particles have formed and there is also still a significant amount of gas remaining, as seen in Figures \ref{fig:frac1} and \ref{fig:frac2}.  Therefore it is an appropriate time to compare how different equations of state affect sink particle formation.  

The most noticeable difference between the isothermal simulation and the simulation with dust heating is the sparseness of sink particles in the simulation with dust heating.  Less particle splitting is also occurring.  When the first sink particles begin to heat the environment, the Jeans mass increases and fragmentation is halted.  Another consequence of dust heating is the lack of definition in the filaments for the simulations with dust heating compared to the isothermal simulation (seen clearly in Fig.\ \ref{fig:xy1}).   The hotter temperatures prevent the filaments from collapsing toward the high density central region.  In Figure \ref{fig:xy2}, there are more filaments in the isothermal simulation (left-hand panel) because fragmentation proceeds unhindered by any increase in temperature, which occurs in the simulation with dust heating (right-hand panel).

\subsection{Density Profile Evolution}\label{sec:ddpe}

\begin{figure}
\plotone{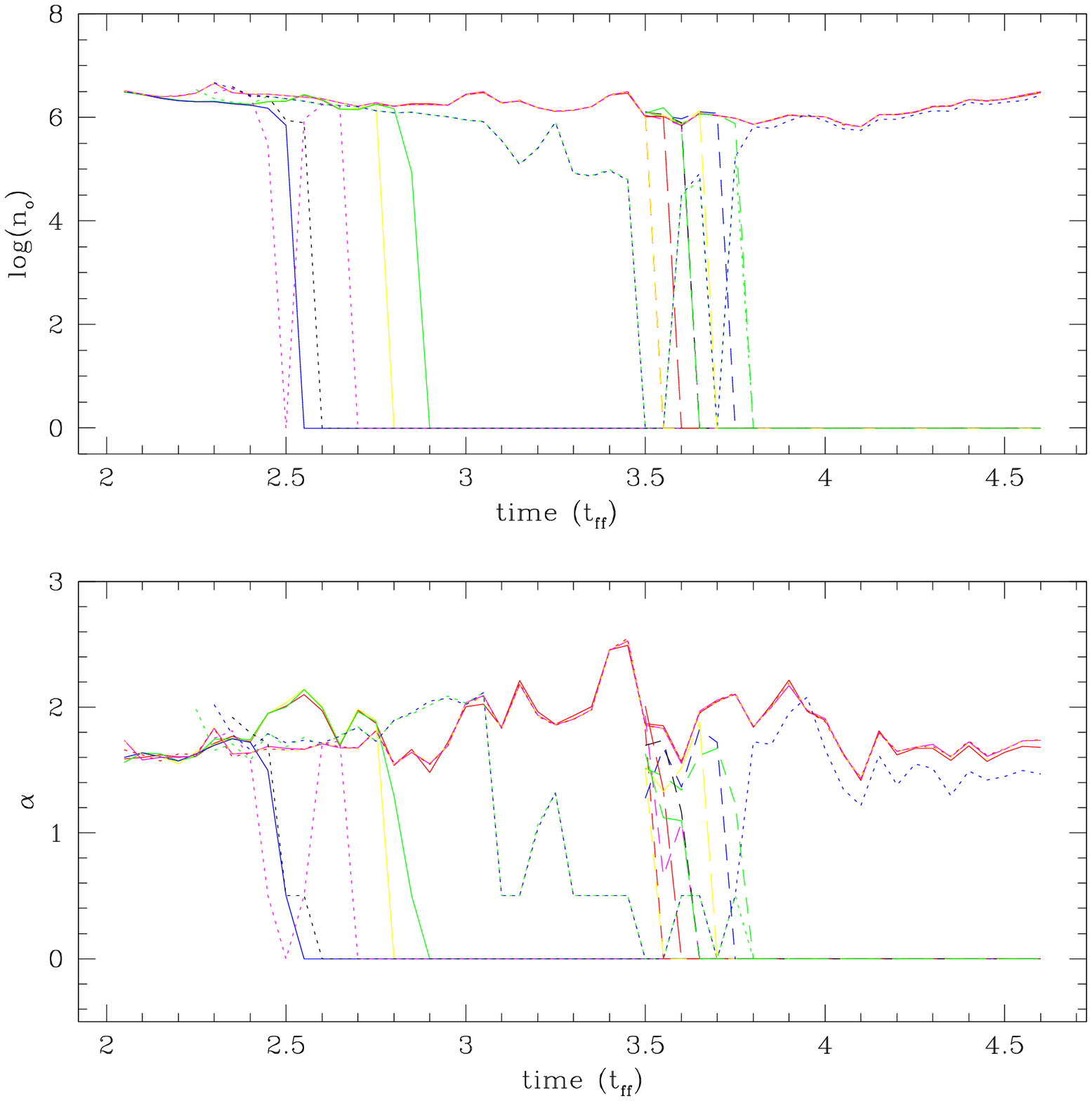}
\caption{Evolution of the density profile parameters for simulation with $N_{\rm gen}=1$, including dust heating.
Each color/line type represents the evolution of a different sink.  Data are sampled every $0.05\tff$.
}
\label{fig:anoevol1}
\end{figure}

\begin{figure}
\plotone{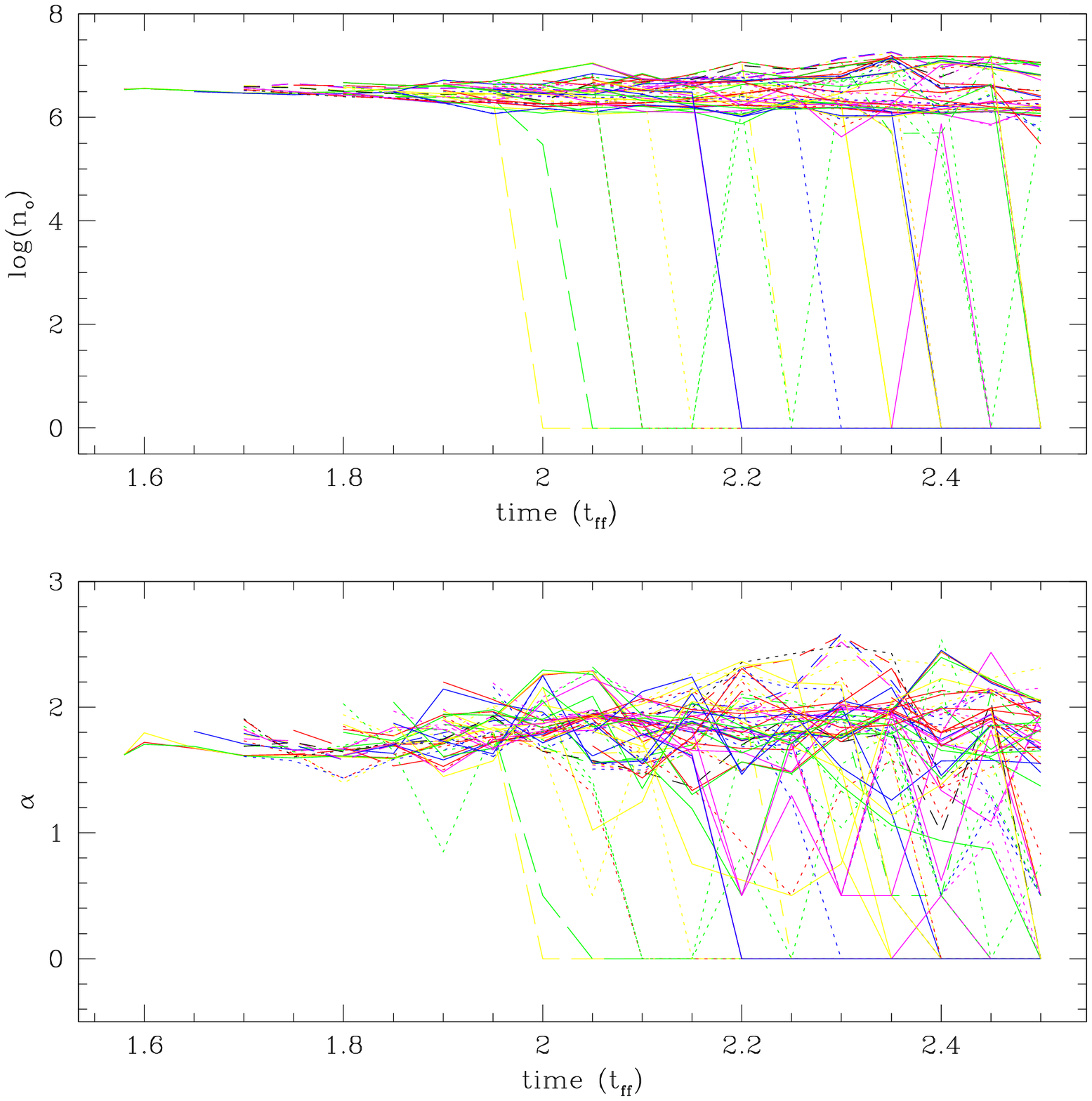}
\caption{Evolution of the density profile parameters for simulation with $N_{\rm gen}=2$, including dust heating.
Each color/line type represents the evolution of a different sink.  Data are sampled every $0.05\tff$.
}
\label{fig:anoevol2}
\end{figure}

\begin{figure}
\plotone{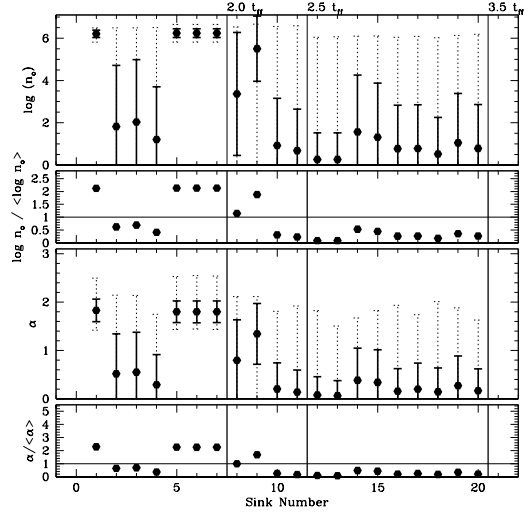}
\caption{Density profile parameters for simulation with $N_{\rm gen}=1$, including dust heating.
Values of density profile parameters, $\alpha$ and $n_o$, 
are shown for the sinks.
The time-averaged values of $\alpha$ and $n_o$ over all sinks are given in the figure.  
Error bars shown with a solid line indicate 
the standard deviation for each individual sink.
Error bars shown with a dotted line indicate the 
minimum and maximum value of $\alpha$ and $n_o$.}
\label{fig:dpe1d}
\end{figure}

\begin{figure}
\plotone{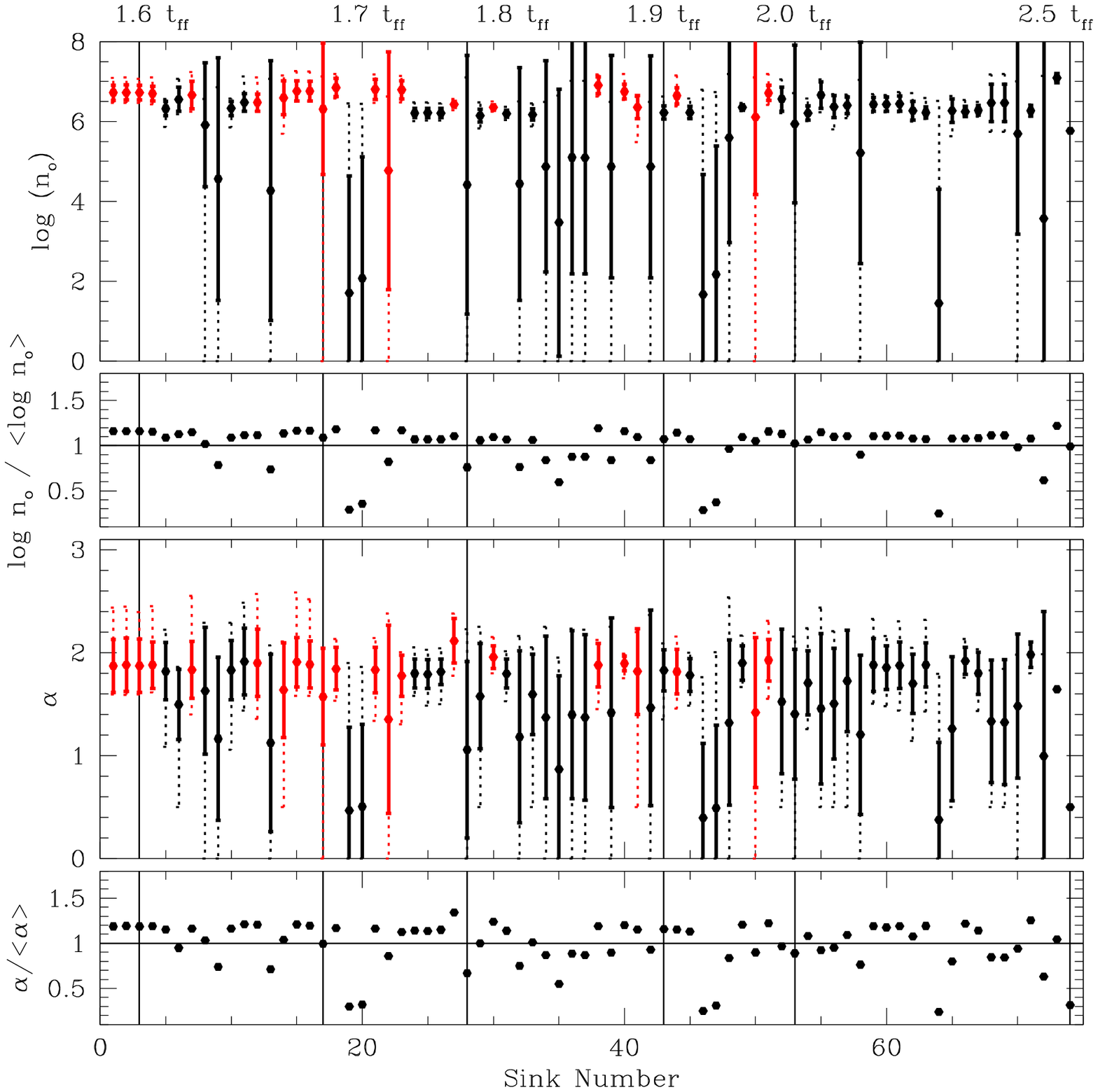}
\caption{Density profile parameters for simulation with $N_{\rm gen}=2$, including dust heating.  Similar to Figure \ref{fig:dpe1d}.  The points that are red are sink particles with final masses greater than $5 \msun$.}
\label{fig:dpe2d}
\end{figure}

At each time step in the simulations with dust heating, we calculate the density profile around each sink particle.  
As the sink particles move around in the simulation box and material accretes, their surrounding density fields change.  
The two parameters that define the density profile, $n_o$ and $\alpha$ are discussed in \S \ref{sec:dens}.  
Figures \ref{fig:anoevol1} and \ref{fig:anoevol2} show the evolution of the density profile for the sink particles formed in the simulations.  
Figures \ref{fig:dpe1d} and \ref{fig:dpe2d} show the average values and dispersions of $\alpha$ and $n_o$ for the individual sink particles.

We calculate the average values of $\alpha$ and $n_o$ using various methods.  The data are sampled from the simulation every $0.05\tff$.  Of the points plotted in Figures \ref{fig:anoevol1} and \ref{fig:anoevol2}, we calculate the average value of (1) all of the points, (2) only points with $\alpha \ne 0$, and (3) only points with $\dot{M} > 10^{-8}$M$_{\odot}$yr$^{-1}$.  We summarize all of the results in Table \ref{tbl:dp}.  The spread is very large when we consider all points.  Ignoring sinks that are not likely to be accreting, i.e., those with $\alpha=0$ and with low accretion rates, we find smaller dispersion in the results.  For case (3), we find $\langle{\alpha}\rangle=1.7 \pm 0.3$, $\langle \textrm{log} (n_o/\rm{cm}^{-3})\rangle  = 6.2 \pm 0.3 $   for simulations with N$_{\rm gen}=1$  and $\langle{\alpha}\rangle=1.7 \pm 0.4$, $\langle \textrm{log} (n_o/\rm{cm}^{-3})\rangle  = 6.5 \pm 0.3 $ for simulations with N$_{\rm gen}=2$.

We can compare our average density profile values to those derived observationally.  Class 0 and Class I cores, which are representative of the earliest stages of isolated low-mass star formation,  have been studied by \citet{shirley} and \citet{young}, respectively.  
\citet{shirley} find $\langle{\alpha}\rangle=1.63 \pm 0.33$ and a typical value of $\alpha=1.8\pm 0.1$ if they ignore two sources with aspherical emission contours.  
\citet{young} find $\langle{\alpha}\rangle=1.6 \pm 0.4$.   
\textit{There is excellent agreement between the average values of $\alpha$ derived for the density profiles around sinks in our simulation and the observed values of $\alpha$ in isolated low-mass star-forming cores.}  
The values of $n_o$ derived from the tables in the papers give $\langle \textrm{log} (n_o/\rm{cm}^{-3})\rangle  = 6.1 \pm 0.2 $ \citep{shirley} and $\langle \textrm{log} (n_o/\rm{cm}^{-3})\rangle  = 5.4 \pm 0.5 $ \citep{young}.  
There is some agreement with our average values; however this comparison may be affected more strongly by other factors such as the masses of the individual cores and the observed intensity to density conversion.

Another interesting feature of the density profile parameters is the relationship between the dispersion of $\alpha$ and $n_o$ and the final mass (shown in Figure \ref{fig:dpe2d}).  We find that objects with high dispersions tend to be the lower mass objects ($M<5\msun$) in our system.  Conversely, objects with the highest mass are more likely to have low dispersions.  (This trend does not appear to apply for sinks formed after $2 \tff$; however, these sinks may not have had enough time to accrete $5\msun$.)

\begin{deluxetable*}{l | c c | c c}
\tablecaption{Density Profile  }
\tablewidth{0pt}
\tablehead{
\colhead{Case  } & \colhead{$N_{\rm gen}=1$} 
& \colhead{}& \colhead{$N_{\rm gen}=2$}  & \colhead{}\\
\colhead{} & \colhead{$\langle{\alpha}\rangle$} & \colhead{$\langle \textrm{log} (n_o/\rm{cm}^{-3})\rangle$ }& \colhead{$\langle{\alpha}\rangle$} & \colhead{$\langle \textrm{log} (n_o/\rm{cm}^{-3})\rangle$ }}
\startdata  
All points   &0.8$\pm$0.9   &  2.9 $\pm$3.1 & 1.6$\pm$0.6  & 5.8$\pm$2.0    \cr       
Points with $\alpha >0$   &1.7$\pm$0.4   &  6.1 $\pm$0.3 & 1.8$\pm$0.4  & 6.5$\pm$0.3    \cr     
Points with $\dot{M} >10^{-8} \msun/$year   &1.7$\pm$0.3   &  6.2 $\pm$0.3 & 1.7$\pm$0.4  & 6.5$\pm$0.3    \cr
\enddata
\label{tbl:dp}
\end{deluxetable*}

\subsection{Temperature and Density Evolution}\label{sec:tempdens}

\begin{deluxetable}{cccccc}
\tablecaption{Temperature (K) Statistics for $N_{\rm gen}=1$ }
\tablewidth{0pt}
\tablehead{
\colhead{time ($\tff$)} & \colhead{mean} & \colhead{sigma} &\colhead{median}
& \colhead{lower} & \colhead{upper} \\
\colhead{} & \colhead{} & \colhead{} &\colhead{}
& \colhead{quartile} & \colhead{quartile}}
\startdata
2.0  & 5.6  &  2.1  & 5.0 & 5.0  &  5.0  \cr 
2.5  & 24   &  3.1  & 22  & 20   &  26 \cr
3.0  & 33   &  13  &  29  & 26   &  34 \cr
3.5  & 36   &  14  &  32  & 30   &  38 \cr
4.0  & 37   &  15  &  32  & 30   &  38   \cr
4.5  & 39   &  20  &  31  & 28   &  40   \cr
\enddata
\label{tbl:tempstats1}
\end{deluxetable}

\begin{deluxetable}{cccccc}
\tablecaption{Temperature (K) Statistics for $N_{\rm gen}=2$ }
\tablewidth{0pt}
\tablehead{
\colhead{time ($\tff$)} & \colhead{mean} & \colhead{sigma} &\colhead{median}
& \colhead{lower} & \colhead{upper} \\
\colhead{} & \colhead{} & \colhead{} &\colhead{}
& \colhead{quartile} & \colhead{quartile} }
\startdata
1.6  &  2.4  &  1.9  &  5.0  &  5.0  &  5.0  \cr 
1.8  &  15   &  6.1  &  12   &  11   &  16  \cr
2.0  &  24   &  9.4  &  21   &  19   &  25  \cr
2.2  &  34   &  15   &  29   &  26   &  34  \cr
2.4  &  47   &   29  &  38   &  33   &  49  \cr
\enddata
\label{tbl:tempstats2}
\end{deluxetable}

\begin{figure}
\epsscale{1.0}
\plotone{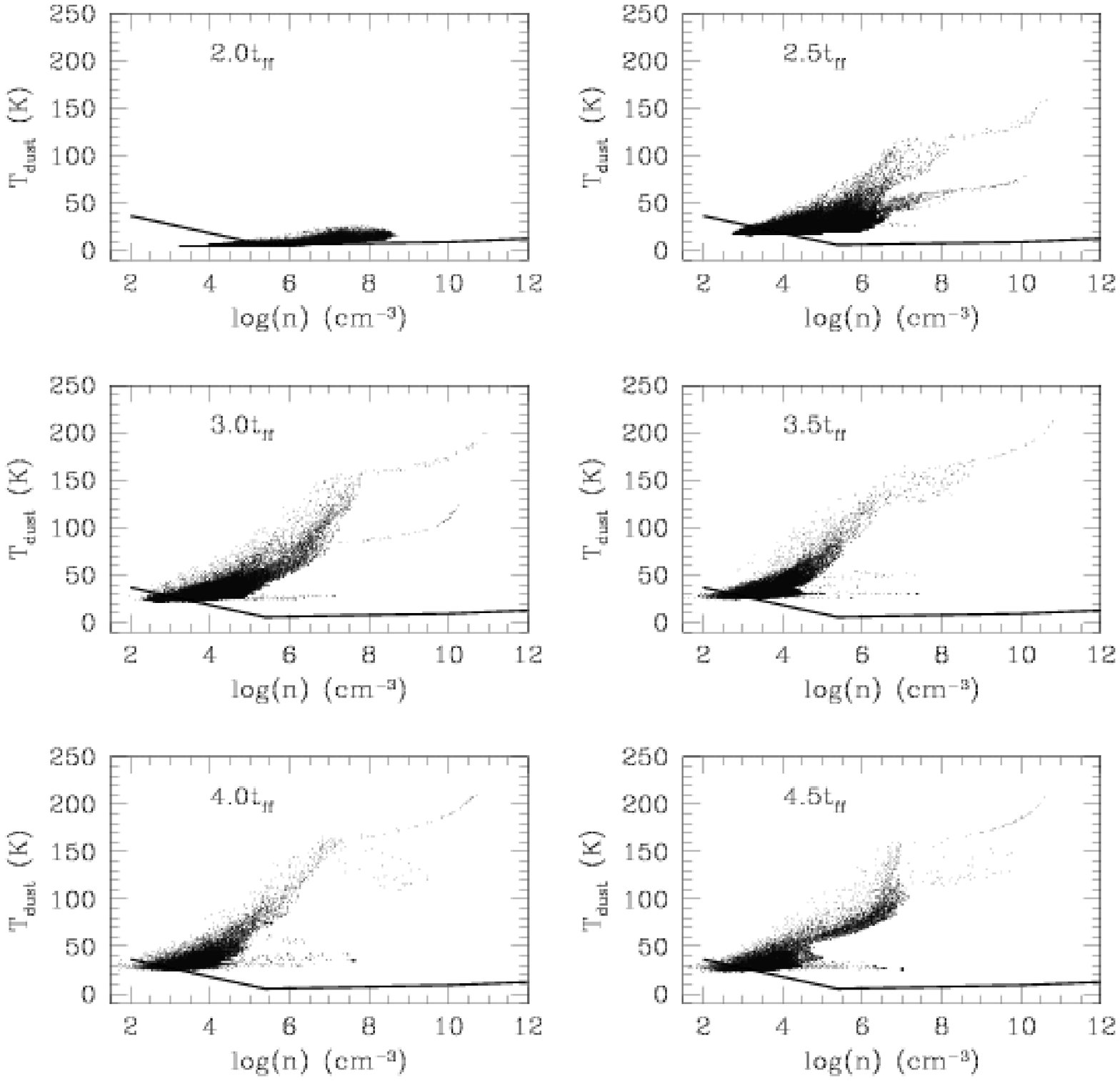}
\caption{Temperature versus density relation of SPH particles as a function of time for simulation with $N_{\rm gen}=1$.  The statistics of the temperature are listed in Table \ref{tbl:tempstats1}.
Solid line shows the equation of state given by \citet{larson}. }
\label{fig:td1}
\end{figure}

\begin{figure}
\plotone{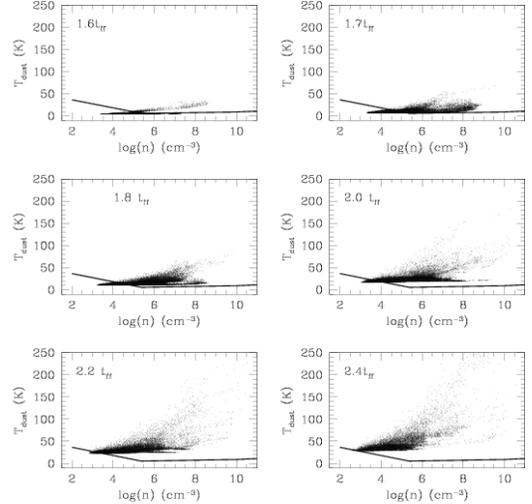}
\caption{Temperature versus density relation of SPH particles as a function of time for simulation with $N_{\rm gen}=2$.   The statistics of the temperature are listed in Table \ref{tbl:tempstats2}.
Solid line shows the equation of state given by \citet{larson}. }
\label{fig:td2}
\end{figure}

Several groups that model clustered star formation assume a simplified equation of state.  For example, \citet{bbb} assume that the gas is isothermal up to a density of $10^{-13}$g/cm$^3$ (or $n=2.6\ee{10}$cm$^{-3}$).  However, this work ignores the effect of stellar heating of the dust and gas.  \citet{larson} gives an equation of state which represents the state of the gas before stars are born and have significant luminosity.   We show the form of this equation as a solid line in Figures \ref{fig:td1} and \ref{fig:td2}.  These figures also show the temperature and density of the gas particles at various times in our simulations.  
The fingers that are seen extending to the right in these plots correspond to the gas particles that are close to a sink particle and therefore have their dust temperature determined by the luminosity of a single sink particle.  
It is clear from Figures \ref{fig:td1} and \ref{fig:td2} that a simple equation of state that describes the behavior of the temperature as a function only of density is insufficient once stars begin to form.  
When this happens, heating depends not only on the local density, but also on non-local effects, such as the distance to a luminous sink and the star formation history.

We give temperature statistics for the two simulations in Tables \ref{tbl:tempstats1} and \ref{tbl:tempstats2}.  For both simulations with dust heating, we find that there is a general trend for denser gas to be hotter.  
Another interesting point is the behavior of the lower quartile which gives the median value of the lower 25\% of the distribution.  This value increases as a function of time indicating that overall the entire region is getting hotter.  
As the sources turn on they slowly heat all of the material in the entire volume.  This ``global warming" is what changes the time evolution and the resulting mass function compared to the isothermal case (see \S \ref{sec:MF}) .

\subsection{Luminosity and Mass Evolution}\label{sec:mlevol}

\begin{figure}
\plotone{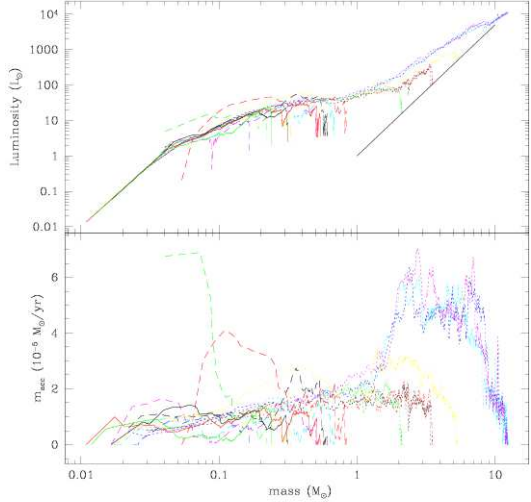}
\caption{Luminosity and mass accretion rate as a function of mass for
simulation with $N_{\rm gen}=1$ and dust heating.  Values of mass, 
mass accretion rate, and luminosity are plotted for all sinks created in 
the simulation with $N_{\rm gen}=1$ and dust heating.  Different colors
and line types correspond to different sinks.  Solid black line in the top panel near the top right corresponds to a slope of 3.7. 
 }
\label{fig:ml1}
\end{figure}

\begin{figure}
\plotone{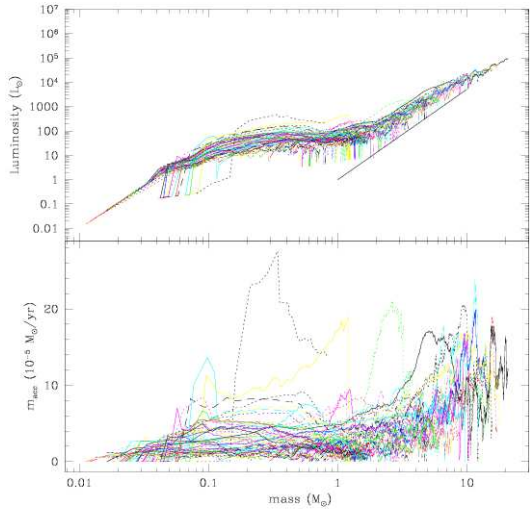}
\caption{Luminosity and mass accretion rate as a function of mass for
simulation with $N_{\rm gen}=2$ and dust heating.  See Figure \ref{fig:ml1}
for details.}
\label{fig:ml2}
\end{figure}

Figures \ref{fig:ml1} and \ref{fig:ml2} show the luminosity and mass accretion rate as a function of mass for the two simulations which include dust heating. (Recall that the mass accretion rate is averaged over a time interval of 5000 years, see \S \ref{sec:lum}.)  
The luminosity as a function of mass follows the same trend in both simulations.  They both also follow the main sequence mass-luminosity relation, $L\propto M^{3.7}$, at high masses with additional luminosity due to accretion.  
Sink particles that suddenly stop accreting can be seen by the lines that drop sharply from the trend of the rest of the sink particles in the plot of mass accretion rate versus mass.  When a sink particle has a low luminosity it can be seen in the mass accretion plot to be in a phase of low accretion. (The same holds for high-luminosity sink particles and high accretion rates.)  Sink particles shown by the red and green dashed  lines in Figure \ref{fig:ml1} at $\sim 0.1 \msun$ demonstrate this effect for low-mass sink particles.

These figures also show that our method of calculating the luminosity using \citet{wt} and the method described in \S \ref{sec:lum} are compatible with one another; the luminosity varies smoothly at 2$\msun$ (the transition mass between the two methods).  There is a slight increase in the slope at 2$\msun$.  This is probably due to the assumption that the stars with $M>2\msun$ have at least a main sequence luminosity as discussed in \S \ref{sec:lum}.  

\begin{figure}
\plotone{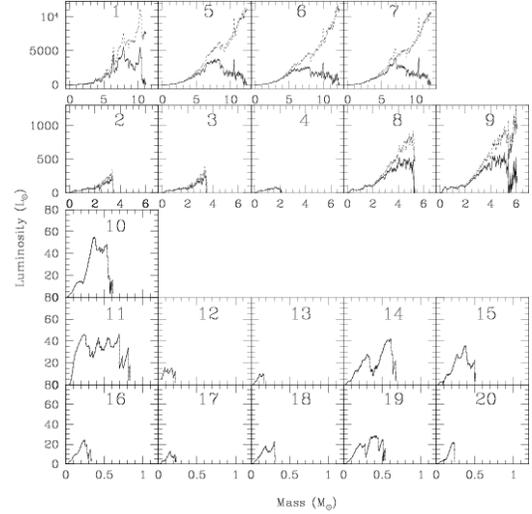}
\caption{Luminosity evolution as a function of final mass for the dust heating simulation with $N_{\rm gen}=2$.  Top row shows the luminosity evolution of sink particles with $M>10\msun$.  Second row shows evolution of sink particles with masses between 1 and 10 $\msun$. Bottom three rows show evolution for sink particles with  $M \le 1\msun$.  Numbers within boxes indicate order of sink formation.  Solid line indicates the luminosity contribution from \citet{wt} luminosity calculation and accretion luminosity, i.e. luminosity from $L=M^{3.7}$ is not included.   Dotted line indicates the total luminosity evolution. When there is no dotted line, there is a negligible contribution from the main sequence luminosity}
\label{fig:lummass}
\end{figure}

In Figure \ref{fig:lummass}, we show the relation between luminosity and mass of the individual sinks in separate panels. This figure demonstrates that high-mass sink particles have a larger fraction of their luminosity due to their main sequence luminosity as they become more and more massive.

There are interesting differences between the two simulations with dust heating.  For the simulation with $N_{\rm gen}=1$ (Fig. \ref{fig:ml1}), the maximum luminosity, mass, and accretion rate are all lower than the maximum values in the simulation with $N_{\rm gen}=2$ (Fig. \ref{fig:ml2}) .  The fact that the $N_{\rm gen}=2$ simulation has larger values of luminosity, mass, and accretion rate is likely to be due to the larger total mass inside the simulation volume (Table \ref{tbl:params}).  The sink particles have more available mass to accrete, which leads to higher sink masses  which in turn affects their mass accretion rate and luminosity.  Another way to interpret this result is that the higher mass in the $N_{\rm gen}=2$ simulation leads to more of the rare, large fluctuations in the density field which allow bigger objects to form.  

\begin{figure}
\plotone{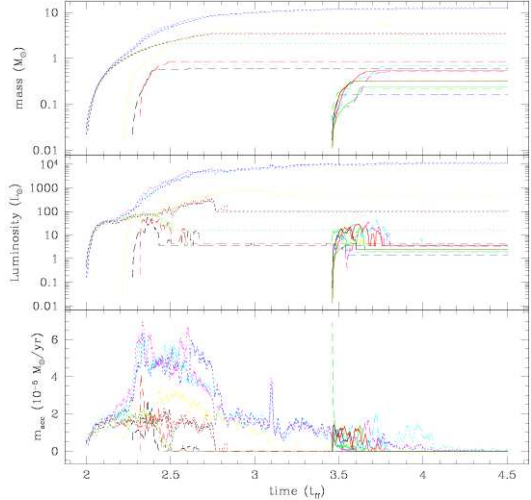}
\caption[Mass, Luminosity, Mass Accretion History of Sinks for simulation with N$_{\rm gen}=1$, Dust Heating.]{Mass, luminosity, mass accretion history of sinks for N$_{\rm gen}=1$.
Time evolution of mass, luminosity, mass accretion rate are shown for all
sinks in simulation with N$_{\rm gen}=1$ and dust heating. Different colors
and line types correspond to different sinks.
 }
\label{fig:mlt1}
\end{figure}

\begin{figure}
\plotone{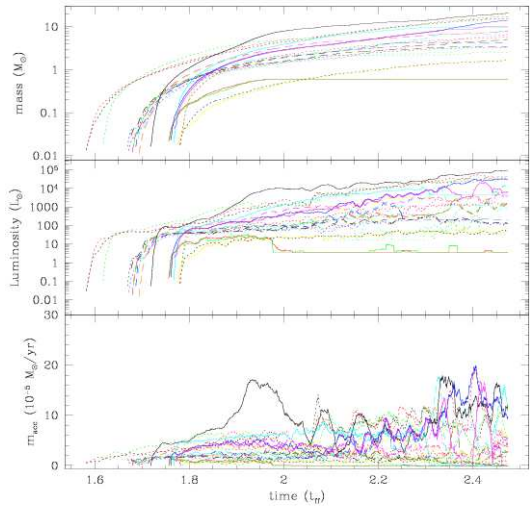}
\caption[Mass, Luminosity, Mass Accretion for simulation with N$_{\rm gen}=2$, A, Dust Heating.]{Mass, luminosity, mass accretion history of first 28 sinks for 
N$_{\rm gen}=2$.
See Figure \ref{fig:mlt1} for details. }
\label{fig:mlt2}
\end{figure}

Figures \ref{fig:mlt1} and \ref{fig:mlt2} show the mass, luminosity, and mass accretion rate as a function of time for the two simulations with dust heating.  If we only compare the time period in our $N_{\rm gen}=1$ simulation which corresponds to the entire runtime of our $N_{\rm gen}=2$ simulation (i.e. $0-2.5 \tff$), then we find similarities and differences between the two simulations.  As Figures 1 and 2 show, at $t=2.5\tff$ a comparable percentage of the mass has been converted into sinks for both simulations, even though the number of sinks formed is  different, 10 sinks for $N_{\rm gen}=1$ and 74 sinks for $N_{\rm gen}=2$.  Comparing Figures \ref{fig:mlt1} and \ref{fig:mlt2} during $0-2.5 \tff$, the formation time between sink particles is greater in the $N_{\rm gen}=1$ simulation.  This is due to the larger sample volume in the $N_{\rm gen}=2$ simulation.

Another interesting difference in the small and large simulations is the time of formation of the most massive object.  Although the most massive object in the $N_{\rm gen}=1$ simulation only reaches 12.4 $\msun$, it is one of the first objects to form in the simulation.  This is not the case for the $N_{\rm gen}=2$ simulation.  For this simulation, the most massive object forms sometime after the formation of the 10th sink particle.  This can be seen in Figure \ref{fig:mlt2}.  The most massive sink particle is represented by the black solid line which forms at time $\sim 1.72 \tff$.

\begin{figure}
\plotone{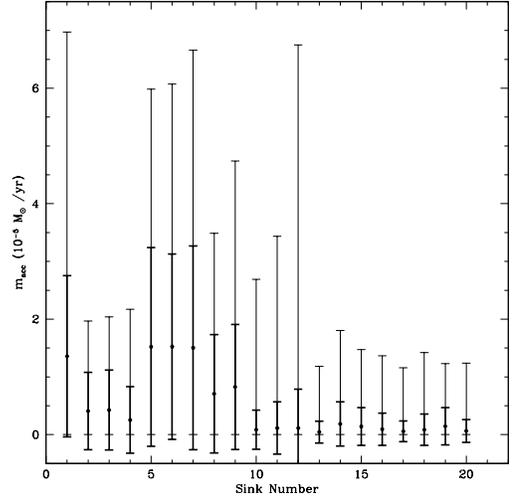}
\caption[Mass accretion rates for $N_{\rm gen}=1$, Dust Heating. ]{Mass accretion rates in simulation with $N_{\rm gen}=1$.  Data was sampled at increments of $\sim 0.02 \tff$.  The average value of mass accretion rate over all sinks  is $\dot{M} = (6.10 \pm 11.47) \ee{-6}\msun$ yr$^{-1}$.  Error bars shown with a thick line indicate the standard deviation for each individual sink.  Error bars shown with a thin line indicate the minimum and maximum value of the mass accretion rate.}
\label{fig:macc1}
\end{figure}

\begin{figure}
\plotone{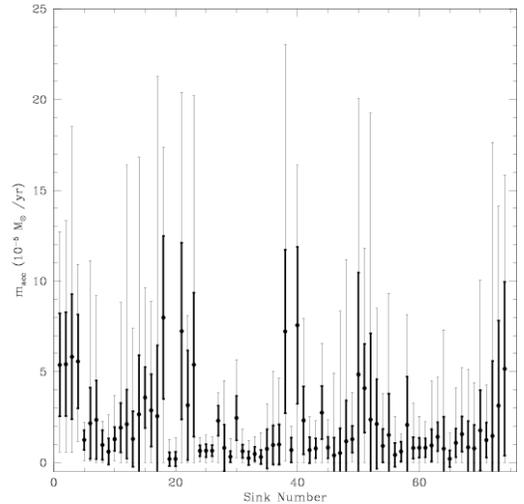}
\caption[Mass accretion rates for $N_{\rm gen}=2$, Dust Heating. ]{Mass accretion rates in simulation with $N_{\rm gen}=2$.  Average value of mass accretion rate over all sinks and all time is $\dot{M} = (2.20 \pm 3.08) \ee{-5}\msun$ yr$^{-1}$. Thick- and thin-lined error bars represent similar values as those discussed in Figure \ref{fig:macc1}'s caption.}
\label{fig:macc2}
\end{figure}

\subsection{Mass Accretion Evolution and Supersonic Accretion}\label{sec:massaccevol}

Figures \ref{fig:macc1} and \ref{fig:macc2} show the average mass accretion rates for each sink particle from the simulations with dust heating.  We find the mass accretion rates to be highly variable.  This was seen previously in \citet{klessen01}.  The time-averaged mass accretion rate over all sink particles is $\dot{M} = (6.10 \pm 11.47) \ee{-6}\msun$ yr$^{-1}$ for the simulation with $N_{\rm gen}=1$ and  $\dot{M} = (2.20 \pm 3.08) \ee{-5}\msun$ yr$^{-1}$ for the simulation with $N_{\rm gen}=2$.  These mass accretion rates are generally higher than those determined observationally from UV, optical, and IR emission excesses in classical T Tauri stars, i.e., 
$\dot{M} \lesssim \eten{-6} \msun$ yr$^{-1}$ (\citealt{hartmann}, \citealt{bc}, \citealt{rw}).  
However, these data come from objects that are more evolved than the objects formed in our simulation and have lost most of their initial envelope.   
They are also low-mass objects ($M \lesssim 1\msun$).  Currently, there are only a few observations of mass accretion rates for young, massive stars.  
\citet{zapata} find a mass accretion rate of $4-7\ee{-2} \msun$ yr$^{-1}$ for W51 North from the observed CN line profile.  The mass of the central object, which could be an O star or a group of B stars, is $\sim 40 \msun$.  \citet{zapata} also list a range of other observed mass accretion rates and masses.  
The rates derived for gas masses with 200-300 $\msun$ and proto-stellar masses with 20-40 $\msun$ are $10^{-4} - 10^{-2} \msun$ yr$^{-1}$.  
These are higher than the high mass accretion rates seen in Figures \ref{fig:macc1}  and \ref{fig:macc2}, but the most massive object that we form is only $20.8 \msun$.

The results shown in Figures \ref{fig:macc1} and \ref{fig:macc2} are plotted against the Sink Number which indicates the order in which the sink particles were formed.  This shows that the order of formation of the sink particles does not have a strong effect on the accretion rate.  For the simulations with $N_{\rm gen}=1$, in which there is less mass, the accretion rate does appear to drop for sink particles formed later.  However, this effect is not seen in the simulation with $N_{\rm gen}=2$, where there are still sink particles with high accretion rates at high Sink Number.

\begin{figure}
\plotone{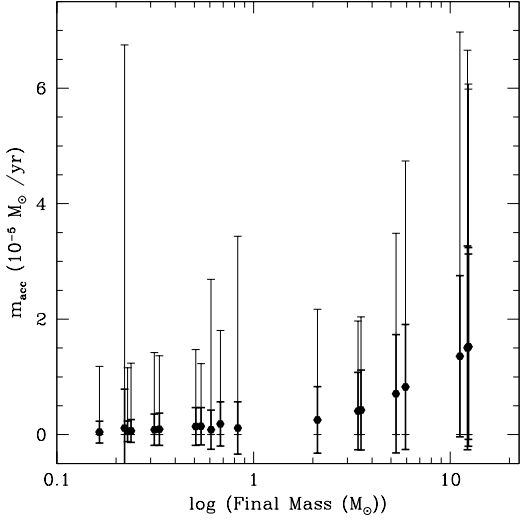}
\caption[Mass accretion rates for $N_{\rm gen}=1$, Dust Heating. ]{Mass accretion rates in simulation with $N_{\rm gen}=1$ as a function of the final sink mass.  Thick- and thin-lined error bars represent similar values as those discussed in Figure \ref{fig:macc1}'s caption.}
\label{fig:maccmass1}
\end{figure}

\begin{figure}
\plotone{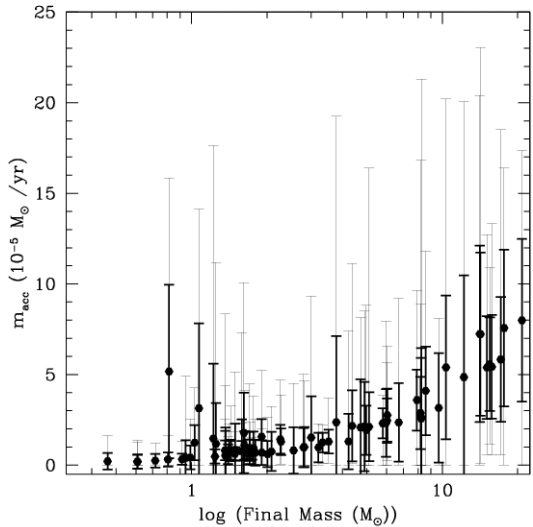}
\caption[Mass accretion rates for $N_{\rm gen}=2$, Dust Heating. ]{Mass accretion rates in simulation with $N_{\rm gen}=2$ as a function of the final sink mass.  Thick- and thin-lined error bars represent similar values as those discussed in the Figure \ref{fig:macc1}'s caption.}
\label{fig:maccmass2}
\end{figure}

In Figures \ref{fig:maccmass1} and \ref{fig:maccmass2} we find a correlation between average accretion rate and the final sink particle mass at masses above $\sim 2 \msun$.   This suggests that high-mass objects are built up with large accretion rates.  This has been suggested by others as a method of overcoming the radiation pressure which could halt the formation of massive young stars (\citealt{YorkeS}, \citealt{KrumholzSci}).  This correlation was either not seen in other similar works because they did not form objects larger than $2 \msun$ \citep{bb05} or the correlation was weak because they formed only a few objects above $2 \msun$ \citep{bate09a, offner}.  Had these earlier works started with larger initial masses, they most likely would have seen a stronger trend once they began forming more massive sink particles.

\begin{figure}
\plotone{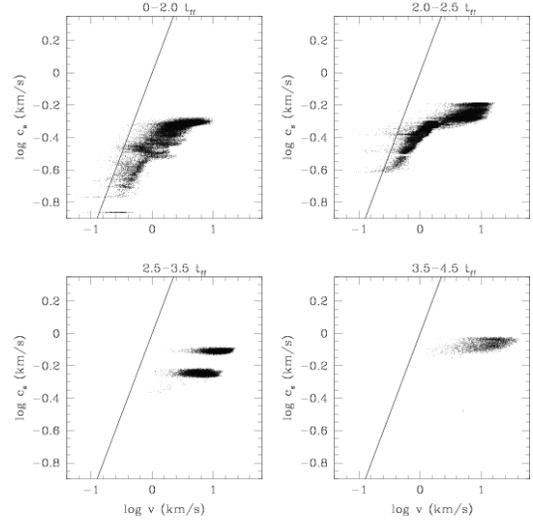}
\caption[Sound speed ($c_s$) versus relative speed ($v$) for $N_{\rm gen}=1$, Dust Heating]{Sound speed ($c_s$) versus relative speed ($v$) of accreted particles plotted for $N_{\rm gen}=1$ simulation with dust heating.  Solid line shows the relation for $
\mathcal M=1$.}
\label{fig:mach1}
\end{figure}

\begin{figure}
\plotone{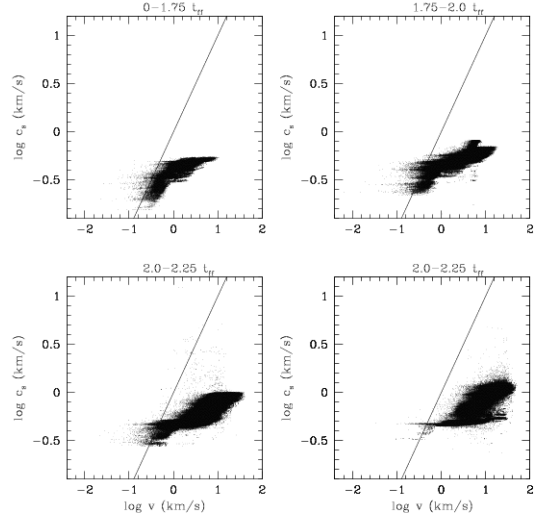}
\caption[Sound speed ($c_s$) versus relative speed ($v$) for $N_{\rm gen}=2$, Dust Heating]{Similar to Figure \ref{fig:mach1} but for $N_{\rm gen}=2$.}
\label{fig:mach2}
\end{figure}

In Figures \ref{fig:mach1} and \ref{fig:mach2}, we show the sound speed versus the speed of the sink particle relative to the accreted particle for the simulations with dust heating at a moment just before the particle is accreted.  Particles are accreting supersonically on average.  The average sound speed of the particles increases over time due to the increase in temperature as seen in Figures \ref{fig:td1}  and \ref{fig:td2}.  
The average Mach number for accreting particles, $\mathcal M$, can be calculated for our simulations.   
For the isothermal simulations,  
$\mathcal M =38.4 \pm 31.5$ ($N_{\rm gen}=1$) and  $\mathcal M = 32.1 \pm 29.6$ ($N_{\rm gen}=2$).    
The values for the simulations with dust heating are much smaller, $\mathcal M=11.8 \pm  7.3$ ($N_{\rm gen}=1$) and $\mathcal M=10.8 \pm 7.4$ ($N_{\rm gen}=2$).  
In all cases, the accretion onto sink particles is supersonic; however, the velocity of accreting particles relative to the sound speed decreases when dust heating is included.

\subsection{Mass Function}\label{sec:MF}

\begin{figure}
\plotone{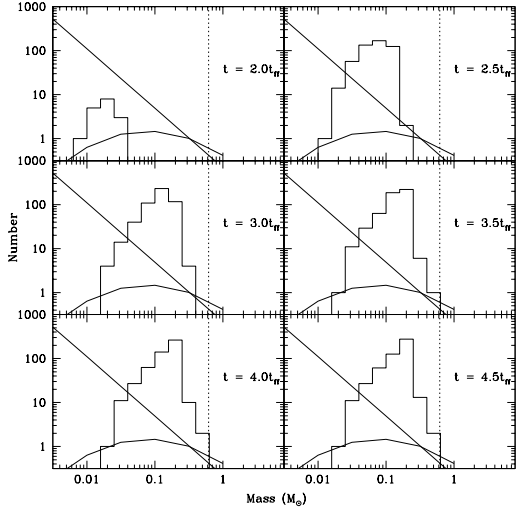}
\caption{Mass histograms for simulation with $N_{\rm gen}=1$ and isothermal equation of state.
Mass function shown at different times listed in top right corner of boxes.
\citet{salpeter} (straight) and \citet{chabrier} (curved) analytic mass function
shown as solid lines.  Dashed line shows initial Jeans mass in the simulation.
}
\label{fig:iso1}
\end{figure}

\begin{figure}
\plotone{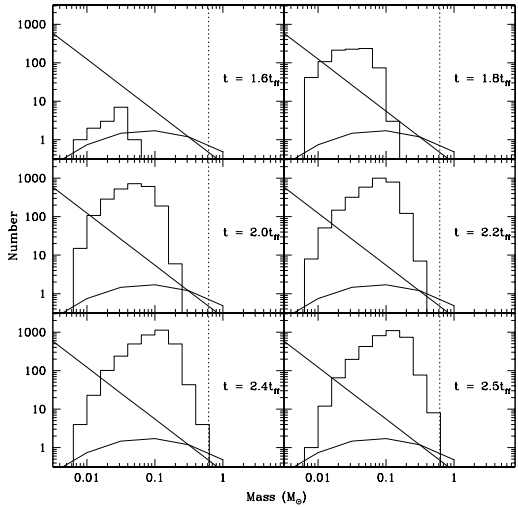}
\caption{Mass histograms for simulation with $N_{\rm gen}=2$ and isothermal equation of state.
See Figure \ref{fig:iso1} for details.}
\label{fig:iso2}
\end{figure}

\begin{figure}
\plotone{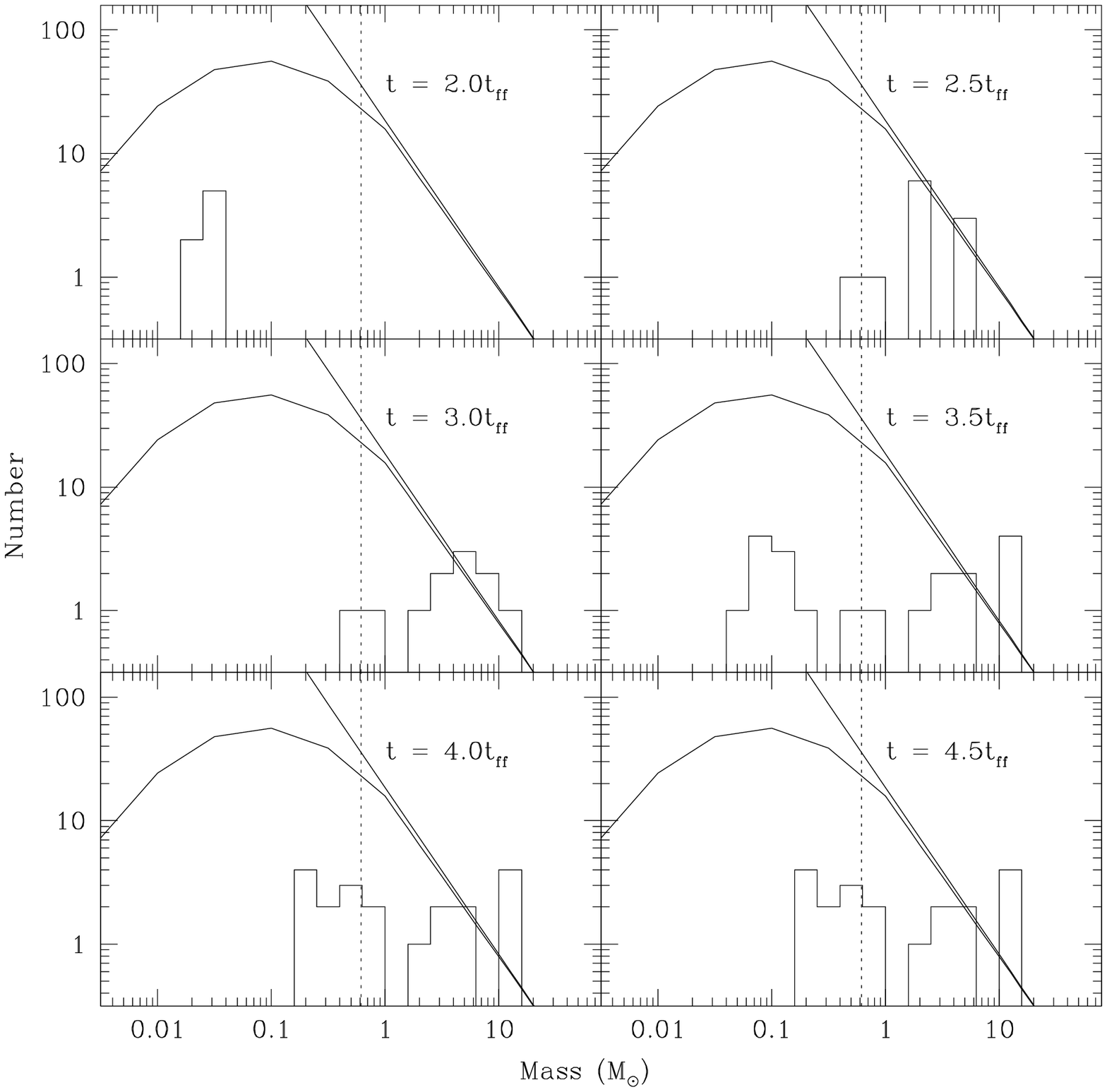}
\caption{Mass Histograms for simulation with $N_{\rm gen}=1$ and dust heating.
See Figure \ref{fig:iso1} for details.}
\label{fig:d1}
\end{figure}

\begin{figure}
\plotone{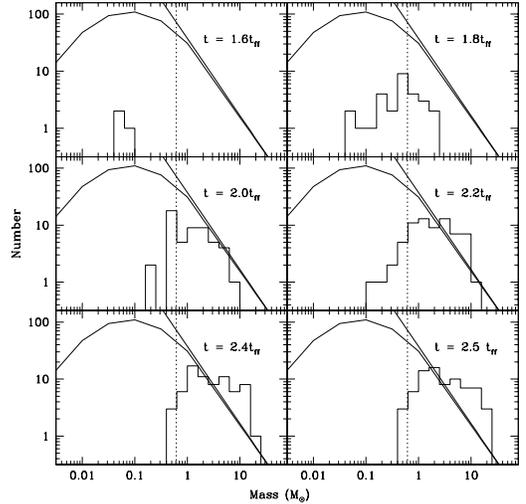}
\caption[Mass histograms for $N_{\rm gen}=2$ and Dust Heating.]
{Mass histograms for simulation with $N_{\rm gen}=2$ and dust heating.
See Figure \ref{fig:iso1} for details.}
\label{fig:d2}
\end{figure}

Figures \ref{fig:iso1}--\ref{fig:d2} show the mass functions of the sink particles in all of our simulations at different times.  Since our simulations with dust heating are isothermal until the sink particles begin heating the surrounding environment, the first sink particles will form at the same time for both the isothermal simulation and simulation with dust heating if the  simulations have the same number of particle splittings (i.e., same values of $N_{\rm gen}$).   The first sink particle for the simulations with $N_{\rm gen}=1$ forms at $t\sim2 \tff$, but for the simulations with $N_{\rm gen}=2$, the first sink particle forms at $t\sim 1.5 \tff$.  Since there is more mass in the simulations with  $N_{\rm gen}=2$, there is a higher probability of forming a sink particle earlier.

The isothermal simulations of MES06 all have the same initial mass but have different values of $N_{\rm gen}$. Hence, an increase in $N_{\rm gen}$ resulted in a higher resolution since the mass per particle was smaller. 
These simulations produced log-normal distributions, with an average value which depended on resolution. As the resolution increased, the mean shifted to lower values.  
In our simulations, discussed in this paper, we use a different approach by keeping the resolution fixed. 
Hence, the total mass of the system increases with the value of $N_{\rm gen}$. 
Based on the results of MES06, we expected the mean value of mass in our simulations to be independent of $N_{\rm gen}$.  As Figures \ref{fig:iso1} and \ref{fig:iso2} show, this is indeed the case.  We find a log-normal distribution with an average value of $\sim 0.1 \msun$ for both of our isothermal simulations.  
In the isothermal simulations, we also find that we are unable to create any objects with masses greater than $\sim 1 \msun$.  
This is the case for many other isothermal simulations and was our motivation for using a dust heating algorithm. 

The mass functions for simulations which include dust heating (Figs. \ref{fig:d1} and \ref{fig:d2}) show that we are able to form massive stars ($M \gtrsim 10\msun$), unlike the isothermal simulations.  However, for the simulation with $N_{\rm gen}=1$ (Fig. \ref{fig:d1}), the distribution is very sparsely sampled.  Adding more mass to the simulation leads to a more well-sampled distribution, even after only $2.5 \tff$ as seen in Figure  \ref{fig:d2} where $N_{\rm gen}=2$.

\begin{figure}
\plotone{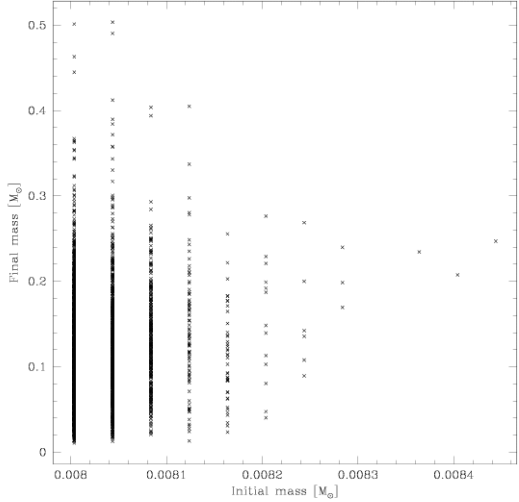}
\caption{Initial versus final mass for isothermal simulation with $N_{\rm gen}=2$. The smallest possible initial sink particle mass is $0.008 \msun$, as discussed in \S \ref{sec:IC}.  This can only increase by the addition of individual gas particles of mass $4 \ee {-5} \msun$.  This explains why sink particles will only form at specific intervals when they form with masses close to the resolution limit of the simulation.}
\label{fig:IF1}
\end{figure}

\begin{figure}
\plotone{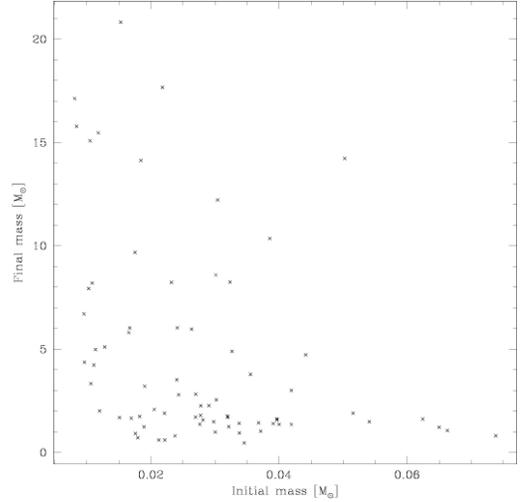}
\caption{Initial versus final mass for Dust Heating simulation with $N_{\rm gen}=2$.}
\label{fig:IF2}
\end{figure}

\begin{figure}
\plotone{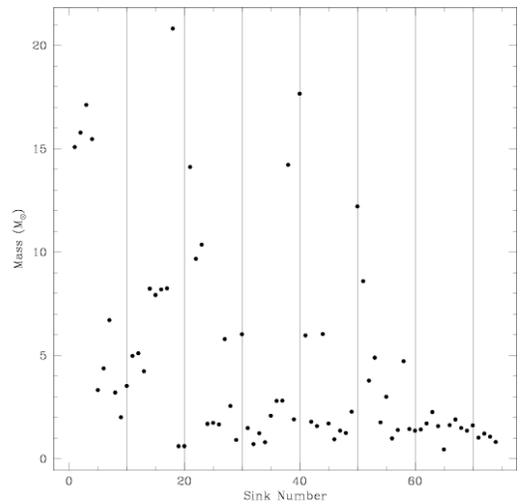}
\caption{Final sink mass as a function of order of formation (Sink Number) for Dust Heating simulation with $N_{\rm gen}=2$.}
\label{fig:massrank}
\end{figure}

In Figures \ref{fig:IF1}--\ref{fig:IF2}, we compare the initial and final masses of the sink particles.  The initial Jeans mass of our simulations is $0.617 \msun$.  If objects were collapsing to the scale of the Jeans mass without fragmentation, then we would expect the average mass of the simulation to be the initial Jeans mass.  For the isothermal simulations, fragmentation is a process which is only halted by the resolution limit of the simulation and therefore many of the sink particles have an initial mass equal to the resolution limit of the simulation and all of the particles have an initial mass which is less than the initial Jeans mass.  Therefore, the average mass in these isothermal simulations is unrelated to the initial Jeans mass (as discussed in MES06).  For the simulations with dust heating, the initial sink masses at the beginning of the simulation are similar to those of the isothermal simulation, as Figures \ref{fig:IF2} and \ref{fig:massrank} show (i.e., objects with high final masses form with low initial masses and early).  However, as the simulation progresses and the temperature heats up, the initial sink masses become larger.   It is important to note that even in this case, the initial sink masses are much smaller than the initial Jeans mass of our simulation, even though the average temperature has increased substantially from the initial 5K.  This seems to indicate that in our simulations the initial Jeans mass does not predict the scale of fragmentation.

Many of the sinks that form in the isothermal simulations are prevented from forming when dust heating is included.  The gas that was destined to form these sinks instead gets eventually accreted by existing sinks, enabling them to reach large masses. As Table \ref{tbl:sim} shows, the final mass that ends up in sinks is comparable in the isothermal and dust simulations, but since the number of sinks is widely different, the masses are different as well.

In Figures \ref{fig:iso1}--\ref{fig:d2}, we also plot two observationally-derived mass functions.  We use the analytic mass functions from \citet{salpeter} and \citet{chabrier} (Disk IMF from Table 1) and normalize them to the maximum sink particle mass in each simulation at the final time.  For the isothermal simulations, the mass function is very different from the two analytic mass functions.  Too many low mass objects are formed given the mass of the most massive object in the simulation.  However, the simulations with dust heating are able to form more massive stars.  For the simulation with $N_{\rm gen}=1$, the mass function appears to under-sample the analytic mass functions at low ($M < 1\msun$) and moderate masses ($\sim 1 \msun$).  The larger simulation ($N_{\rm gen}=2$) shows more promising results.  The mass function is better sampled.  The slope at the high masses is very similar to the Salpeter slope.  However under-sampling at low and intermediate masses is still present.  

At the same time, $t = 2.5 \tff$, both simulations ($N_{\rm gen}=1$ and 2), have a similar mass distribution with the same range of masses.  As the simulation with $N_{\rm gen}=1$ advances in time, its most massive objects gain mass and more low mass objects are formed.  We cannot probe past $t = 2.5 \tff$ in our $N_{\rm gen}=2$ simulation because the maximum allowable mass would be exceeded (see \S \ref{sec:timescales}).  However, as we mention in \S \ref{sec:SPH}, we assume sink particles do not undergo further fragmentation once they have formed.  This may not be the case.  If further fragmentation does occur within some of our sink particles, these events may populate the low mass end of the distribution in Figure \ref{fig:d2}.

\section{Discussion}\label{sec:discuss}

The investigation performed in this work was a study of the effect of dust heating on the star formation process.  This work is not intended to explain the complete star formation process.  As we discuss in \S \ref{sec:results}, an obvious missing factor in our work is turbulence.  Another important factor is magnetic fields \citep{price08, price09}.  In the following two sections we compare our work to other similar work and then discuss future improvements.

\subsection{Comparison to Previous Work}
In this section, we compare our work and results to those of \citet{krumholz}, \citet{bate09r}, and  \citet{offner}.  The most obvious difference between our work and theirs is size.  Our largest simulation ($N_{\rm gen}=2$) models a box with  $M = 671 \msun$ and $L=0.984$pc.  It exceeds in size the largest simulation of \citet{krumholz} which models $200 \msun$ in a sphere of radius $0.1$pc or \citet{offner} with a box of mass $185 \msun$ and $L=0.65$pc.  Because of the differences in scale of our simulations, it is somewhat difficult to compare global properties since these other simulations do  not form the large number of objects that we do.  However, we all find that including the effect of radiative transfer drastically decreases the number of objects formed.  (This can be understood from a Jeans mass argument.  If the gas is hotter, in this case, due to radiation, then the amount of fragmentation will decrease, thereby reducing the number of objects formed.)  Some of the other results that we find are only hinted at in  \citet{bate09r} and  \citet{offner}, namely the change in the mass function and the correlation of accretion rate and mass.

The highest number of objects formed in \citet{krumholz} is 7.  Therefore, they do not attempt to produce a mass function.  The same is true of \citet{offner} in which they form 15 objects.  \citet{bate09r} forms 17 objects and produces a mass function which he compares to observed mass functions.  However, with such a small number of objects and a maximum stellar mass  less than $2 \msun$, it is difficult to draw conclusions.  In our largest simulation with dust heating, we form 74 objects.  Our mass function samples masses up to $20\msun$, and while accretion is still occurring in our simulation, we can see more conclusively that including dust heating encourages the formation of massive stars, while inhibiting the fragmentation that leads to an overabundance of low-mass objects in simulations that do not include dust heating.  

\citet{offner} finds a slight trend of increasing accretion rate with mass.  In \citet{bate09a} (which uses similar simulation parameters as \citet{bate09r}), there is also a hint of a trend of accretion rate with mass.  The trend in our simulation begins at $\sim 2 \msun$ and if objects larger than $10 \msun$ are ignored then the trend is difficult to see in our smaller dust-heating simulation (see Fig.\ \ref{fig:maccmass1}) which only formed 20 objects, similar to the number of objects formed in \citet{bate09r} and \citet{offner}.  However, if we look at our larger simulation with 74 objects, then the trend is obvious (see Fig.\ \ref{fig:maccmass2}).

As discussed in \S\ref{sec:intro},  \citet{krumholz}, \citet{bate09r}, and  \citet{offner} all use FLD to calculate the dust temperature in their simulations.  Using this method in the optically thick regions of \citet{krumholz} is probably valid.  However, for the cases of \citet{bate09r} and  \citet{offner}, they study lower density regions than \citet{krumholz} and it is unclear if the FLD approximation is still valid.   All three also ignore the wavelength dependence of the dust opacity when calculating the radiation field.  
Our method also makes approximations, but of a different nature.  We assume that the material is spherically distributed around the sink particles.  Based on the figures in 
\citet{krumholz}, \citet{bate09r}, and  \citet{offner} this is clearly not always the case.   Both methods use approximations, therefore, it is difficult to say which is a  ``better'' method.    Since there is currently no realistic method of using three-dimensional, wavelength-dependent radiative transfer, it may be the case that the best alternative is a combination of our two methods.   At early stages when the gas is less dense and the density distribution around the sink particles is roughly spherical, our method may be more appropriate.  However, as the density increases and disks begin to form around stars, the FLD method may be more appropriate.   
It is important to note that, despite the different radiative transfer methods used with a variety of approximations, the main conclusion of our work and the works of \citet{krumholz}, \citet{bate09r}, and  \citet{offner} is the same: heating severely inhibits the fragmentation of the gas and promotes the formation of massive stars.

One main difference between the work of \citet{bate09r} and our work (and the work of \citet{krumholz} and \citet{offner}) is their approximation regarding the source of radiation.  \citet{bate09r}  ignores much of the accretion luminosity and does not consider nuclear burning. The first assumption may not be valid during the early stages of star formation because accretion may be at its highest level then.  
\citet{krumholz} and  \citet{offner}  do not ignore the effect of protostellar heating, therefore their work does not suffer from the problem of missing radiation as in  \citet{bate09r}.   They both include a stellar model to calculate the accretion and intrinsic luminosity from the protostars in their simulations.  As discussed in the following section, \S \ref{sec:future}, we plan to use a similar approach using a stellar evolution model in our future work in order to improve our calculation of the stellar luminosity.

Since \citet{krumholz} and  \citet{offner} include a stellar model for their sink particles, we can compare the stellar properties from our simulations. If we compare our smaller simulation ($N_{\rm gen}=1$) to their simulations, we find that our luminosities are comparable.  The accretion rate in \citet{krumholz} and  \citet{offner} is found to be highly variable, similar to our results.  
We can only qualitatively investigate the density distribution around the objects formed in the simulations of \citet{krumholz}, \citet{bate09r}, and  \citet{offner} via their simulation figures.   From these it is not clear how to compare them with the spherically-averaged density distributions that we have calculated around our sink particles.   However, they investigate disk properties around some of their objects which we have not done.

Another main difference between our work and that of \citet{bate09r}, \citet{krumholz}, and \citet{offner} is that our work does not include compressional heating or viscous heating that is included in the FLD method.  \citet{offner} find that compressional heating does not dominate the heating, but rather stellar heating is most important throughout most of their simulation.  However, before stars have formed, heating via viscous dissipation is dominant.   Therefore, the effects that we have ignored are most likely to change how fragmentation proceeds in the beginning of the simulation and then how the first stars form.  Taken together, our work and the work of \citet{bate09r} show the importance of different heating mechanisms throughout the star formation process.

One minor difference between our work and that of \citet{offner} is their inclusion of turbulent driving throughout their simulation.  Since our simulations are on different scales, it is not clear how this affects their results compared to ours.  
Another minor difference between our work and others' is that we do not allow sink particles to merge.  \citet{krumholz}, \citet{bate09r}, and  \citet{offner} all allow this to occur;  however for \citet{bate09r} this never happens.  Therefore, it is not clear if it is necessary to include this process in our simulations.

\subsection{Future Improvements}\label{sec:future}

Our method is an approximation of what we believe to be one of the most important effects in the very early stages of star formation, namely dust heating via young stars.  We have attempted to model this stage as accurately as possible, yet there are areas which we believe can be improved in future work.  

\textbf{(1)}  We do not account for all possible sources of radiation in our simulation.  We have attempted to account for heating of the gas during the collapse using the models of \citet{wt}.  However, there may be some extra contractional heating that we do not include when our sink particles have a mass less than $0.01 \msun$.

Besides including these effects, we can also improve the radiative transfer method discussed in this paper.  Our method of calculating the dust temperature assumes spherical symmetry, yet images of young star forming regions show three-dimensional morphology.  A more advanced method of calculating the dust temperature, at the current level of computational power, is impractical.  Even though we have already made approximations to reality, we have had to extrapolate in our dust temperature look-up table.   Our extrapolation was necessitated by the fact that we could not know a priori the range of envelope profiles that would be created in our simulation.  This is discussed in more detail in \S \ref{sec:tdust}.  The extrapolation occurred in regions which we believe were fairly well-understood.  We can address this in future work with an expanded interpolation table.   

\textbf{(2)} We currently assume that the dust and gas temperature are equal.  While this may be the case for very dense regions studied by \citet{krumholz}, this is not always the case for our model.  We will address this issue in a forthcoming paper.

\textbf{(3)} To calculate the luminosity, we have had to interpolate in a small table of mass and mass accretion rate \citep{wt}, which introduced uncertainties in our luminosity calculation.  Our method of calculating luminosity could be improved because we currently assume that luminosity varies smoothly with mass and mass accretion rate independent of past history.  We plan to address this in future work using an advanced stellar evolution code.  

\textbf{(4)} As mentioned before, fragmentation may be occurring within sink particles.  We do not believe that this will strongly influence the temperature and thus the fragmentation of the gas that has not yet formed into sinks.  
Consider the case of a high-mass sink that fragments (within the sink radius) 
into two equal or two highly dissimilar mass bodies. 
In the first case, the intrinsic luminosities from the two 
equal-mass stars would sum to less than the intrinsic luminosity of a single 
object with the same total mass because of 
the steep dependence of intrinsic luminosity on mass.  
However, the accretion luminosity would remain the same in both cases.  As seen in Figure \ref{fig:lummass}, the accretion luminosity is a substantial contributor to the luminosity of objects with $M>10\msun$.  
Therefore, we do not expect the luminosity to change substantially in the first case.  In the second case, the luminosity missing from the most massive sink due to fragmentation would be negligible.  However, fragmentation within a sink could affect the mass function of our simulation.  
This fragmentation could either decrease the mass of the most massive object in our simulation or it could populate the low-mass end of the mass function.  
This issue could be explored in future work by increasing the resolution of our simulation.

\section{Summary and Conclusion}\label{sec:con}

We have investigated the effect of the heating of dust via luminosity sources in a clustered star formation simulation.  We compare the results of  isothermal simulations to simulations that include dust heating.  Including the effect of dust heating drastically reduces the number of objects formed (by more than an order of magnitude).  We find that the density profiles of the envelopes surrounding the sinks/cores formed in our simulations with heating are comparable to those found around isolated, low-mass star-forming cores.  This brings up the question of how similar density profiles can be formed in such different accretion environments.  Another interesting result of our simulations is that the accretion of mass onto the sinks/cores is found to be highly variable, in contrast to what is theorized for isolated, low-mass star-forming cores.   We also find a strong correlation between the average accretion rate and the final mass for objects with $M>2 \msun$.  This fact may provide a clue to how massive stars form.   We also analyze the final mass function of our simulations.  We find that we are able to reproduce the results of MES06 for our isothermal simulations, i.e., a log-normal distribution centered at very low masses ($\sim 0.1 \msun$) with no objects with masses greater than $1 \msun$.   The mass functions produced by our simulations that include dust heating show that we are able to produce massive stars ($M \gtrsim 10\msun$).
However, we do see a dearth of objects at low and intermediate masses. 
 This may be due to the extreme heating by the dust and the lack of cooling physics.  In our next paper we plan to relax the assumption of dust-gas collisional coupling at all densities.  
We will include the complete energetics algorithm described in \citet{urban}, which includes molecular cooling and cosmic-ray heating, in future simulations similar to the ones discussed in this paper. 

\acknowledgments
This work benefited from stimulating discussions with S. Doty.  All calculations were performed at the Laboratoire d'astrophysique num\'erique, Universit\'e Laval.  We are pleased to acknowledge the support of NASA Grants NAG5-10826, and NAG5-13271.  We would also like to thank the Canada Research Chair program (H.M.), NSERC (H.M.), NSF Grant AST 0607793 (N.E.), and the NASA GSRP Fellowship Program (A.U.) for providing support for this work.  Part of A. Urban's contribution to the research described in this paper was carried out at the Jet Propulsion Laboratory, California Institute of Technology, under a contract with the National Aeronautics and Space Administration. 
 (c) 2009.  All rights reserved.

\bibliographystyle{apj} 
\bibliography{diss}  

\end{document}